\documentclass[epsfig,color]{article}
\pdfoutput=1

\usepackage{bm}
\usepackage{graphicx}

\newcommand{\beq}{\begin{equation}}
\newcommand{\eeq}{\end{equation}}
\newcommand{\bmat}{\begin{displaymath}}
\newcommand{\emat}{\end{displaymath}}

\begin{document}

\title{Quench dynamics of the 
$2d$ XY model} 
\author{Asja Jeli\'c$^{1}$ and Leticia F. Cugliandolo$^{2}$\\
$^{1}${\small 
Universit\'e Paris-Sud, Laboratoire de Physique Th\'eorique,
CNRS UMR 8627,}\\
{\small  B\^atiment 210, Orsay F-91405, France} \\
$^{2}${\small Universit\'e Pierre et Marie Curie - Paris 6,} \\
{\small Laboratoire de Physique Th\'eorique et Hautes Energies, CNRS UMR 7589,}
\\
{\small 4, Place Jussieu, Tour 13, 5\`eme \'etage, 75252 Paris Cedex 05, France}
}
\date{\today}

\maketitle

\maketitle

\begin{abstract}
We investigate the out of equilibrium dynamics of the two-dimensional
XY model when cooled across the Berezinskii-Kosterlitz-Thouless (BKT)
phase transition using different protocols.  We focus on the evolution
of the growing correlation length and the density of topological
defects (vortices).  By using Monte Carlo simulations we first
determine the time and temperature dependence of the growing
correlation length after an infinitely rapid quench from above the
transition temperature to the quasi-long range order region.  The
functional form is consistent with a logarithmic correction to the
diffusive law and it serves to validate dynamic scaling in this
problem. This analysis clarifies the different dynamic roles played by
bound and free vortices.  We then revisit the Kibble-Zurek mechanism
in thermal phase transitions in which the disordered state is plagued
with topological defects. We provide a theory of quenching rate
dependence in systems with the BKT-type transition that goes beyond
the equilibrium scaling arguments. Finally, we discuss the
implications of our results to a host of physical systems
with vortex excitations including planar ferromagnets and liquid crystals
as well as  the Ginzburg-Landau approach to bidimensional freely 
decaying turbulence.
\end{abstract}

\section{Introduction}

The out of equilibrium dynamics of systems with continuous symmetries
annealed or quenched from the symmetric to the symmetry broken phase
are a subject of study in different branches of physics.
Liquid-crystals quenched across a phase transition into their ordered
phase present a large variety of topological defects that diffuse,
interact and eventually annihilate~\cite{liquid-crystals}. Vector
ferromagents cooled across their Curie temperature are other
condensed-matter examples with a variety of dynamic topological
defects~\cite{planar-magnets}. Cosmology provides a set of interesting
models in which similar dynamics occur~\cite{cosmology}. Still, the
decay of vortex density in freely decaying turbulence has also been of
interest~\cite{Tabeling}. In all these areas, there is
considerable interest in the dynamics of defects once the phase
transition has been crossed.

Of special interest is the planar time-dependent Ginzburg Landau model
with SO(2) symmetry or its lattice counterpart the $2d$ XY model. A
static phase transition occurs at a finite critical temperature,
$T_{KT}$, between a high-$T$ disordered paramagnet and a low-$T$
ferromagnetic phase with quasi-long range order.  When the model is
quenched from the high to the low temperature phase the continuous
symmetry of the disordered phase is broken to the one of the ordered
phase. In these systems the defects are singular vortices that carry
topological charge and have logarithmic interactions.  In the high-$T$
phase free vortices proliferate while in the low-$T$ phase vortices
bind in pairs. Physical realizations are bidimensional planar
ferromagnets~\cite{planar-magnets}, superconducting
films~\cite{Bishop80}, Josephson-junction arrays~\cite{Beasley},
especially tailored nematic liquid crystals~\cite{Pargellis94}, and
toy models for bidimensional turbulence~\cite{Tabeling}.  The
exponential singularity may also describe the critical properties of
superfluid helium films~\cite{Helium}.

Phase ordering kinetics following an infinitely rapid quench through a
thermal critical point have been extensively studied in the
statistical physics literature~\cite{coarsening-reviews}.  In
condensed matter applications the time spent close to the critical
point is typically much shorter than the time spent far from it and,
therefore, the infinitely fast quench approximation is justified. A
few studies where cooling rate dependencies have been taken into
account are~\cite{Stinchcombe,FisherHusecooling,Yoshino,Krapivsky}.
The study of cooling-rate dependencies in defect dynamics is, instead,
central in cosmology.  The goal in this case is to estimate the
density of defects, $\rho$, left over in the universe after it went
through a number of potential (second-order) phase transitions.  The
importance lies on the fact that, initially, these were thought to act
as seeds for matter clustering. This scenario seems to be excluded by
observation data now but, still, the interest in predicting the
density of topological defects remains due to other possible effects
of these objects~\cite{Kibble76}.  In the late 80s Zurek derived a
quantitative prediction for $\rho$ that is based on critical scaling
in the disordered phase close to a second order phase transition.  The
proposal is based upon the hypothesis whereby the defect dynamics
should be negligible below the transition. In consequence, $\rho$ is
assumed to remain fixed to the value it takes at the control parameter
at which the system falls out of equilibrium in the disordered phase.
In his articles Zurek proposed to check these predictions in condensed
matter systems with the same symmetry properties as the cosmological
models~\cite{Zurek}.  A vast experimental~\cite{KZ-exp,Rivers-exp} and
numerical~\cite{KZ-num} activity followed with variable results
summarized in~\cite{Kibble07}.

The fact that the annihilation of defects can be neglected in the out
of equilibrium regime was questioned in~\cite{Biroli}.  In this paper,
a scaling argument that includes the mechanism of defect annihilation
in the low-temperature phase for systems with second-order phase
transitions and dissipative dynamics was proposed. A
scaling of the density of defects with cooling rate {\it
  and} time spent in the out of equilibrium evolution was derived.
The predictions were checked numerically with Monte Carlo dynamics of
the $2d$ Ising model, a paradigmatic system with discrete symmetry
breaking. Similar ideas about the importance of the dynamics below the 
transition were stressed in~\cite{Antunes-Rivers}.

The treatment of all cases mentioned in the previous paragraph is classical.
Recently, pushed by the advent of powerful experimental techniques in 
cold atom systems, the extension of the Kibble-Zurek mechanism to quantum 
{\it isolated} systems was developed~\cite{Zoller,Polkovnikov,Dziarmaga}.
These claims have been critically revisited in~\cite{Santoro,Dutta}, as 
reviewed in~\cite{Silva}.

In this paper we examine the density of vortices left over after going
through the Berezinskii-Kosterlitz-Thouless classical phase transition~\cite{Berezinskii,Kosterlitz}
with a finite speed.  We revisit the Kibble-Zurek mechanism and we
provide a theory of quenching rate dependence in systems with
continuous symmetry and the BKT-type of `infinite-order' phase
transition that goes beyond the equilibrium scaling arguments.  We
test it with a Monte Carlo numerical study of the $2d$ XY model on a
square lattice. We work in the canonical setting, in the sense that
the system is coupled to an equilibrium environment that allows for
energy dissipation.  As a control parameter driving the phase
transition we use the temperature of the heat-bath. For concreteness
we use linear cooling procedures, characterized by a single parameter,
the cooling rate.  A previous numerical study of this problem with a
different method that ignores the effects of free-vortices appeared
in~\cite{Chu}.  We shall discuss the results in this paper and
confront them to ours in the body of this work.

The organization of the paper is the following. 
In Sect.~\ref{sec:model} we recall the definitions of the $2d$ XY model 
and the time-dependent Ginzburg-Landau equation and we present 
the Monte Carlo method used. In Sect.~\ref{sec:BKT} we shortly review the 
BKT scenario. Section~\ref{sec:quenching} is devoted to the analysis of infinitely
rapid quenches. In Sect.~\ref{sec:annealing} we present our results on 
annealing procedures. Finally, in Sect.~\ref{sec:conclusions} we give our conclusions.

\section{Model and method}
\label{sec:model}

In this section we recall the definition of the $2d$ XY model and 
we describe the Monte Carlo method that we use in our simulations of the 
lattice model. We also explain 
the 
corresponding time-dependent Ginzburg-Landau stochastic equation.

\subsection{The $2d$ XY model}

The two-dimensional XY model is defined by the Hamiltonian
\begin{equation}
 H = -J \sum_{\left\langle i,j \right\rangle} {\vec s}_i \cdot {\vec s}_j,
\end{equation}
with ferromagnetic exchange coupling, $J>0$. The spins are classical
variables constrained to live on a plane. ${\vec s}_i^2=1$, and the
sum runs over nearest neighbours on a square lattice of linear size
$L$. It is convenient to use a parametrization in which the spin
vector is represented by the angle it forms with a chosen axis, ${\vec
  s}_i=e^{\imath \theta_i}$; with it the Hamiltonian becomes
\begin{equation}
 H = -J \sum_{\left\langle i,j \right\rangle} \cos \left(\theta_i - \theta _j
\right)\, .
\end{equation}
Above the critical temperature $T_{KT}$, unbound 
positively and negatively charged vortices are
present and the system is disordered. Below
$T_{KT}$ vortex-antivortex pairs bind and the system is critical 
 in a sense that we
shall make more precise below. The critical temperature
is found to be $k_B T_{KT}\simeq 0.89 J$~\cite{Kosterlitz}.
We measure $T$ in units of $J/k_B$ in the following.

\subsection{Dynamics}

Stochastic dynamics, mimicking the coupling of the classical spins to
a thermal bath, can be attributed in different ways. In this Section
we define the Monte Carlo procedure used in our simulations. We focus
on non-conserved order-parameter dynamics (model A) in which the
magnetization is not conserved. We also briefly recall the
time-dependent Ginzburg-Landau approach.

\subsubsection{Monte Carlo dynamics}
\label{subsec:MonteCarlo}

We performed a Monte Carlo study of the $2d$ XY model on a square
lattice of linear size $L$ with periodic boundary conditions. 
The Monte Carlo algorithm consists in updating the angular variable $\theta_i$
associated to a randomly chosen site $i$ to a new value $\theta_i'$ randomly
chosen in the interval $\left[-\pi,\pi\right]$, with probability 
$p=\min\left\{1,e^{-\beta \Delta E}\right\}$, where $\Delta E$ is the energy
variation between the two configurations. 
Equilibrium data are for relatively small system sizes, $L=16$ to $L=50$, while 
out of equilibrium ones are for larger samples, $L=100$ to $L=400$. We briefly discuss
finite size effects when presenting the simulation data. The results are an
average of $5$ to $10000$  independent runs each (depending on the quantity calculated) and one unit of time 
is an attempted move of every spin.

In infinitely rapid quenches from infinite temperature we took a fully
random initial condition in which a random number $\theta_i$ in the
interval $\left[-\pi,\pi\right]$ is associated to site $i$. We explain
the implementation of more sophisticated cooling procedures in
Sect.~\ref{sec:annealing}.

The focus of our study is the evolution of the number of vortices.  We
determine it numerically with two types of measurements.  On the one
hand dynamic scaling implies that the density of vortices, $\rho_v$,
should depend on the tyical growing lengh, $\xi$, as $\rho_v \simeq
\xi^{-2}$. We shall use several determinations of $\xi$ that include 
the exponential decay of the two-point correlation function, $C[\xi_{c}(t,T),t]=const.$, 
and the second moment of $C(r,t)$, that is defined as follows~\cite{Hasenbusch}. 
From the total magnetization  $\vec{M} (t)= \sum_i\vec{s}_i(t)$
and the magnetic susceptibility 
\begin{equation}
 \chi (t) = \frac{1}{L^2} \vec{M}^2 (t),
\end{equation}
the second moment correlation length on a lattice of size $L^2$ is defined by
\begin{equation}
 \xi_{2}^2(t) = \frac{1}{\left(2\sin(\pi/L)\right)^2} \left( \frac{\chi(t)}{F(t)} -1\right),
\end{equation}
where
\begin{equation}
 F(t) = \frac{1}{L^2} \sum_{ij}  \left\langle \vec{s}_{i}(t)\cdot\vec{s}_{j}(t)\right\rangle
           \cos\left( 2\pi(j_{x}-i_{x}) \right).
\end{equation}

 In addition, we determine the number of vortices $N_v$
directly from the configuration of the system by counting the
plaquettes with non-zero vorticity. The vorticity is determined as the
integer winding number $n$, so that the phase difference
$\theta_{ij}\equiv \theta_{i}-\theta_{j}$ around the plaquette is
equal to $\sum \theta_{ij}=2\pi n$.  The number of vortices $N_v$ is
then obtained as the number of plaquettes around which the phase
$\theta_{i}$ rotates through $\pm 2\pi$, taking care that the phase
difference $\theta_{ij}$ is restricted to the interval
$\left[-\pi,\pi\right]$, as are the phases $\theta_{i}$.  The density
of vortices is then $\rho_v=N_v/L^2$.

In the figures we will not report the statistical errors because they are very small
for the quantities we consider (smaller than dot dimensions in the various plots). 
For example, for
the largest system size ($L=400$) for which the fluctuations
are larger, the variance of the number of vortices is at most 3 to 4\%.

\subsubsection{Time-dependent Ginzburg-Landau equation}

A field theory version of this problem, better suited for analytic calculations
is given by 
the time-dependent Ginzburg-Landau equation that determines the 
stochastic evolution of the coarse-grained two-component field $\vec \phi$: 
\begin{equation}
\frac{\partial \vec \phi(\vec r,t)}{\partial t}  = -\Gamma \frac{\delta H}{\delta \vec \phi(\vec r,t)}
\label{eq:time-dep-GL}
\end{equation}
with $\Gamma$ the kinetic coefficient. The `free-energy' is
\begin{equation}
H= \int d^2r \left[ \frac{\rho_s}{2} (\vec \nabla \vec \phi(\vec r))^2
+ \frac{u}{4}  (\phi^2(\vec r) -1)^2 \right] 
\end{equation}
with $\rho_s$ the stiffness coefficient or elastic constant and $u>0$
 the strength of the non-linearity that drives the vector field 
to take unit modulus.  Noise
can be added to Eq.~(\ref{eq:time-dep-GL}) in the form of an
additional stochastic term in its right-hand-side with Gaussian
statistics and, typically, delta-correlated, $\langle\zeta_a(\vec r,t)
\zeta_b(\vec r', t') \rangle=2 k_BT \Gamma \delta_{ab} \delta(\vec
r-\vec r') \delta(t-t')$, with $a,b=1,2$.  The analytic predictions on
the time-dependent growing length to be discussed in
Sect.~\ref{subsec:analytic-xi} have been obtained using this
approach~\cite{Huse,Bray94}.

\section{The BKT equilibrium phase transition}
\label{sec:BKT}

 We summarize 
the static properties of the $2d$ XY model as figured out by Berezinskii~\cite{Berezinskii} 
and Kosterlitz and Thouless~\cite{Kosterlitz} (the BKT picture).

In the thermodynamic limit $2d$ XY models display no conventional
long-range order (the magnetization vanishes at all temperatures due
to spin-wave excitations) but topological order below a critical
temperature $T_{KT}$.  The physics is dominated by two types of
excitations: harmonic spin waves and vortices. The former are
responsible for destroying conventional long-range order while the
latter are responsible for the phase transition.  The
Berezinskii-Kosterlitz-Thouless (BKT) phase transition at $T_{KT}$ is
characterized by a change in the behaviour of the equilibrium spatial
correlation function $C(r) =\left.\langle \vec s_i \vec s_j
\rangle\right|_{|\vec r_i - \vec r_j|=r} $ where the angular brackets
denote an average over the equilibrium measure.  The high-$T$ phase is
disordered, the correlation decays exponentially,
\begin{equation}
C(r) \simeq e^{-r/\xi_{eq}}
\; , 
\end{equation}
and there is a finite density of
free vortices. 
Close and above $T_{KT}$ the correlation length diverges exponentially
\begin{equation}
\xi_{eq} \simeq a_\xi \  e^{b_\xi [(T-T_{KT})/T_{KT}]^{-\nu}}
\label{eq:equil-corr-length}
\end{equation}
with $\nu=1/2$ and $b_\xi$ a non-universal constant that typically
takes a value of order one (on a square lattice $b_\xi \sim
1.5$)~\cite{Berezinskii,Kosterlitz}. Thermodynamic quantities are
smooth accross the transition.  The low-$T$ phase has quasi-long-range
order, 
\begin{equation}
C(r) \simeq r^{-d+2-\eta(T)} = r^{-\eta(T)}
\; ,
\label{eq:lowT-corr}
\end{equation}
it is critical in the sense that the equilibrium correlation length
diverges, $\xi_{eq}\to\infty$, and it is characterized by the
$T$-dependent exponent $\eta(T)$~\cite{Kosterlitz} that decreases upon
decreasing temperature from $\eta(T_{KT})=1/4$ to $\eta(T\to 0) = T/(2\pi J)$. 
Bound vortex-antivortex pairs populate the low-temperature ordered
phase coexiting with spin-waves.  The transition is interpreted by
using an analogy with the Coulomb gas in which the system has a
low-temperature dielectric phase (with charges, in the $2d$ XY case
vortices, bound into dipoles) and a high-temperature plasma or
conducting phase with free charges (in the $2d$ XY model
vortices)~\cite{Berezinskii,Kosterlitz}. In short, at $T_{KT}$ pairs 
dissociate. 

The BKT singularity yields the best fit to data generated with
large-scale Monte Carlo simulations and short-time dynamics.
The data tend to confirm the analytical values of the exponents $\nu$
and $\eta(T)$ and the critical temperature was estimated to
$T_{KT}\simeq 0.89$~\cite{Wolff89,Gupta92,Luo97,Luo98,Ozeki03}. In finite 
size systems the effective transition temperature behaves as 
$T_{KT}+O(\ln^{-2}L)$~\cite{Bramwell}.
Monte Carlo values of the correlation length $\xi_{eq}(T)$ obtained with
lattices with linear size $L=512$ are $\xi_{eq}(1.25)\simeq 3.8$,
$\xi_{eq}(1.04)\simeq 18.7$, and $\xi_{eq}(0.98) \simeq 70$~\cite{Gupta92}.
We shall discuss our own data, obtained with much 
smaller system sizes, in Fig.~\ref{fig:eq-corr-length}. 

\section{Relaxation after a quench}
\label{sec:quenching}

The nonequilibrium behavior of the $2d$ XY model following an
instantaneous quench below $T_{KT}$ has been studied theoretically as
well as experimentally~\cite{Pargellis94,Nagaya92}.  In this Section
we recall the time-dependence of the growing correlation-length. We go
beyond this by-now well-established aspect of the dynamics with a
careful analysis of the temperature-dependence that will be of use in
the rest of the article. We present a detailed analysis of the 
vortex-anti-vortex time-dependent structure.

\subsection{Analytic prediction for the growing length}
\label{subsec:analytic-xi}

In this Section we set the origin of time to be the instant at which
the initial condition is let evolve in the new parameter conditions.
At all $T\leq T_{KT}$ a system prepared in an out of equilibrium
initial state approaches equilibrium through a coarsening process in
which local -- critical -- equilibrium is established over a length
scale $\xi(t,T)$. The space-time correlation $C(r,t) = \left. \langle
\vec s_i(t) \vec s_j(t)\rangle\right|_{|\vec r_i-\vec r_j|=r}$ is
governed by the scaling form
\begin{equation}
C(r,t) \simeq r^{-\eta(T)} f\left(\frac{r}{\xi(t,T)}\right)
\; .
\end{equation}
$\eta(T)$ is the static critical exponent. The scaling function
depends on the initial conditions. We focus on high-temperature,
disordered ones. $f(0)=1$ so that the equilibrium result
(\ref{eq:lowT-corr}) is recovered for all $a \ll r\ll \xi(t,T)$, $a$ being 
a microscopic cut-off length, and,
in particular, in the infinite long time limit in which $\xi(t,T) \to
\infty$.  In the opposite limit $r\gg \xi(t,T)$ the system remains
disordered, as in the initial condition, and this is ensured by an $f(x)$
that rapidly falls-off for $x\gg 1$.

The relaxation of an out of equilibrium state is due to two processes:
the dynamics of Goldstone modes and the annihilation and pairing of
vortices. One can tune the importance of the vortex contribution by
choosing the initial condition. For example, a completely ordered
configuration is free of vortices while a high temperature
configuration has a finite density of free vortices.  After taking a
fully ordered initial condition to $0<T<T_{KT}$, no free vortices are
generated by thermal fluctuations and the system orders in the new
conditions by growing a length $\xi(t,T) \propto [\lambda_0(T)
t]^{1/2}$~\cite{Cukupa,Berthier}. Instead, the relaxation of an
initial condition with free vortices in the critical low-temperature
phase is far more interesting.  Whether dynamic scaling holds and,
in the affirmative, which
is the time-dependent growth-law was a subject of
debate for some time. The following scenario is well-established now.

The numerical integration of the zero-temperature Ginzburg-Landau
equation using fully random initial conditions (with, on average, a
density of vortices of $1/3$) first suggested a power-law for $\xi(t,T)$
with a small but measurable deviation from the diffusive law
$t^{1/2}$~\cite{Goldenfeld90,Huber}. This regime was considered to be a
transient fully determined by the vortex annihilation dynamics.  Yurke
{\it et al.}~\cite{Huse} and Bray {\it et al.}~\cite{Bray94} built
upon these results and gave an argument for
\begin{equation}
\xi(t,T) \simeq \left\{\lambda (T) \ \frac{t}{\ln [t/t_0(T)]}\right\}^{1/z}
\qquad\qquad z=2
\label{eq:growth-law}
\end{equation}
in quenches from equilibrium at $T\geq T_{KT}$ to $T=0$, see below. 
For the sake of completeness, we wrote this expression inserting the 
dynamic exponents $z$ ($=2$ in this case). 
The extension to finite working $T$
was worked out in~\cite{Bray00}.
This law was confirmed by extensive zero-temperature numerical
simulations in~\cite{Rojas99,Bongsoo}. The parameter $\lambda(T)$ 
and the microscopic cut-off
$t_0(T)$ are non-universal in the sense that they depend on the procedure
used to measure $\xi$. 

Let us briefly reproduce here the argument that leads to (\ref{eq:growth-law})
and extend it later to infer the relaxation time.
Within the Ginzburg-Landau model (\ref{eq:time-dep-GL}), 
a field configuration, describing a single free vortex $\vec\phi=\vec r/r$
 has an energy $E_v = \pi\rho_s \ln(L/a)$, where $L$ and $a$
 are the system size and microscopic cut-off ({\it e.g.}, the lattice spacing). 
 In the many-vortex situation, where $\xi(t)$ represents the typical spacing 
 between vortices and anti-vortices one can derive the following 
 expression for the typical speed of a vortex~\cite{Huse,Bray00}
 \begin{equation}
 \frac{d\xi(t)}{dt} \simeq \frac{\rho_s\Gamma}{\xi(t)\ln [\xi(t)/a]}
 \end{equation}
 from the zero temperature time-dependent 
 Ginzburg-Landau equation~(\ref{eq:time-dep-GL}). 
The solution is Eq.~(\ref{eq:growth-law}) with
  $t_0 \simeq a^2/\rho_s\Gamma$.  
We assume, following~\cite{Bray00}, that this equation holds at finite $T$, and even above 
$T_{KT}$ where, though, the growing length saturates to the equilibrium correlation 
length at sufficiently long times. Matching the growing length with the equilibrium law by 
$d\xi/dt|_{\xi_{eq}} \simeq \xi_{eq}/\tau_{eq}$ above $T_{KT}$ one finds 
\begin{equation}
\tau_{eq} \simeq \xi_{eq}^{z} \ln (\xi_{eq}/a)
\; . 
\label{eq:taueq}
\end{equation}
An exponential divergence of $\tau_{eq}$ with the distance to criticality 
is recovered by substituting 
$\xi_{eq}$ with the expression in~(\ref{eq:equil-corr-length}):
\begin{equation}
\tau_{eq} \simeq a_\tau \ e^{b_\xi z[(T-T_{KT})/T_{KT}]^{-\nu}} 
\
\left(\frac{T-T_{KT}}{T_{KT}}\right)^{-\nu}
\label{taueq-Tdep}
\; .
\end{equation}
The same behavior is estimated in~\cite{Jonsson} with a different argument.

\subsection{Numerical measurements of the growing length}

Although different determinations of the growing length 
should converge, asymptotically to the same and unique
law, some are more convenient than others at finite times. 
Comparison of different prescriptions (from scaling of the 
space-time correlation, the density of topological defects, 
and the structure factor) were discussed in~\cite{Rojas99,Bongsoo}. We 
revisit this issue with the purpose of clarifying the different
r\^ole played by bound and free vortices and establishing the 
temperature dependence of the correlation length.

\begin{figure}[h]
\hspace{1cm} (a) \hspace{5.5cm} (b)
\begin{center}
\includegraphics[width=0.49\textwidth]{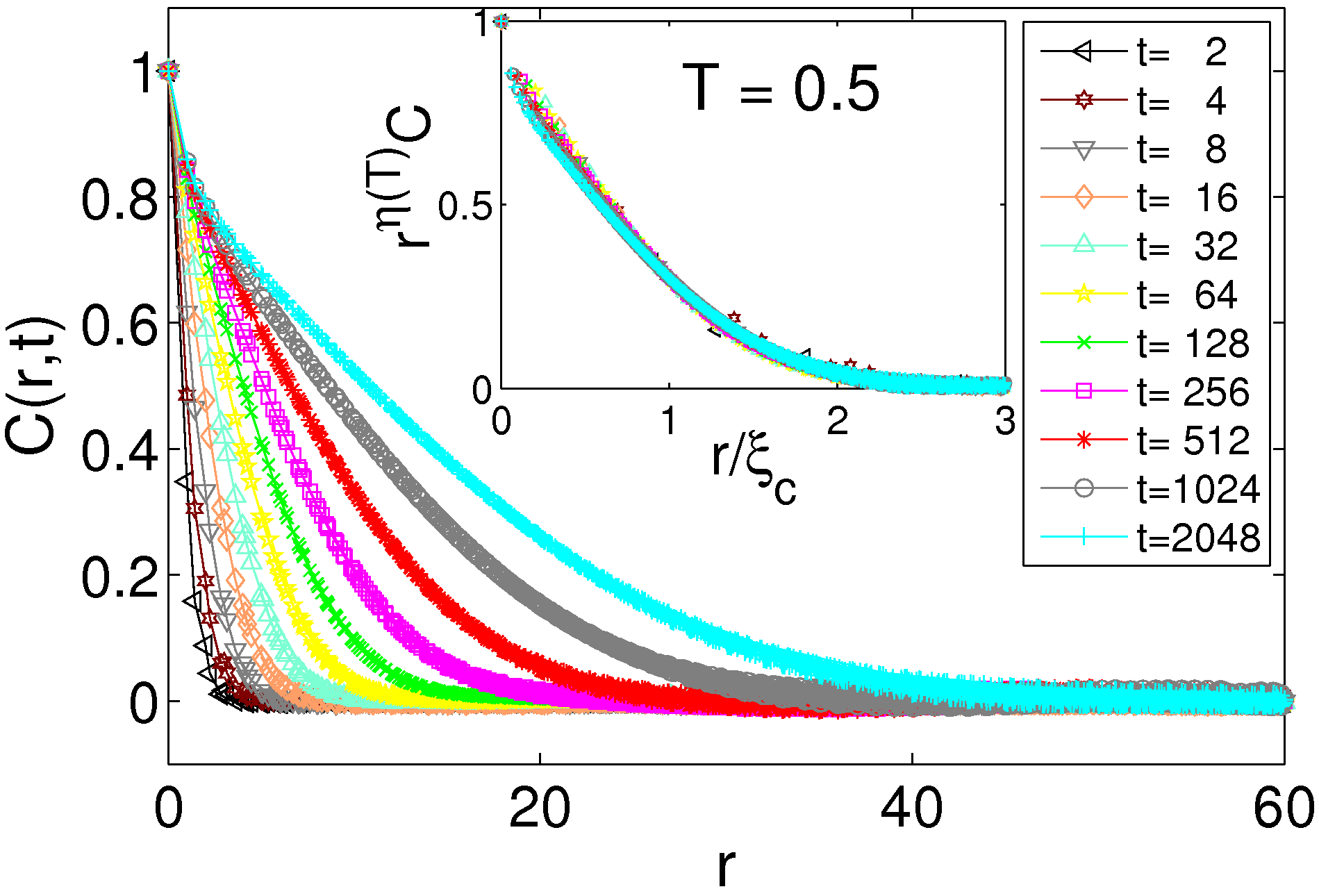}
\includegraphics[width=0.49\textwidth]{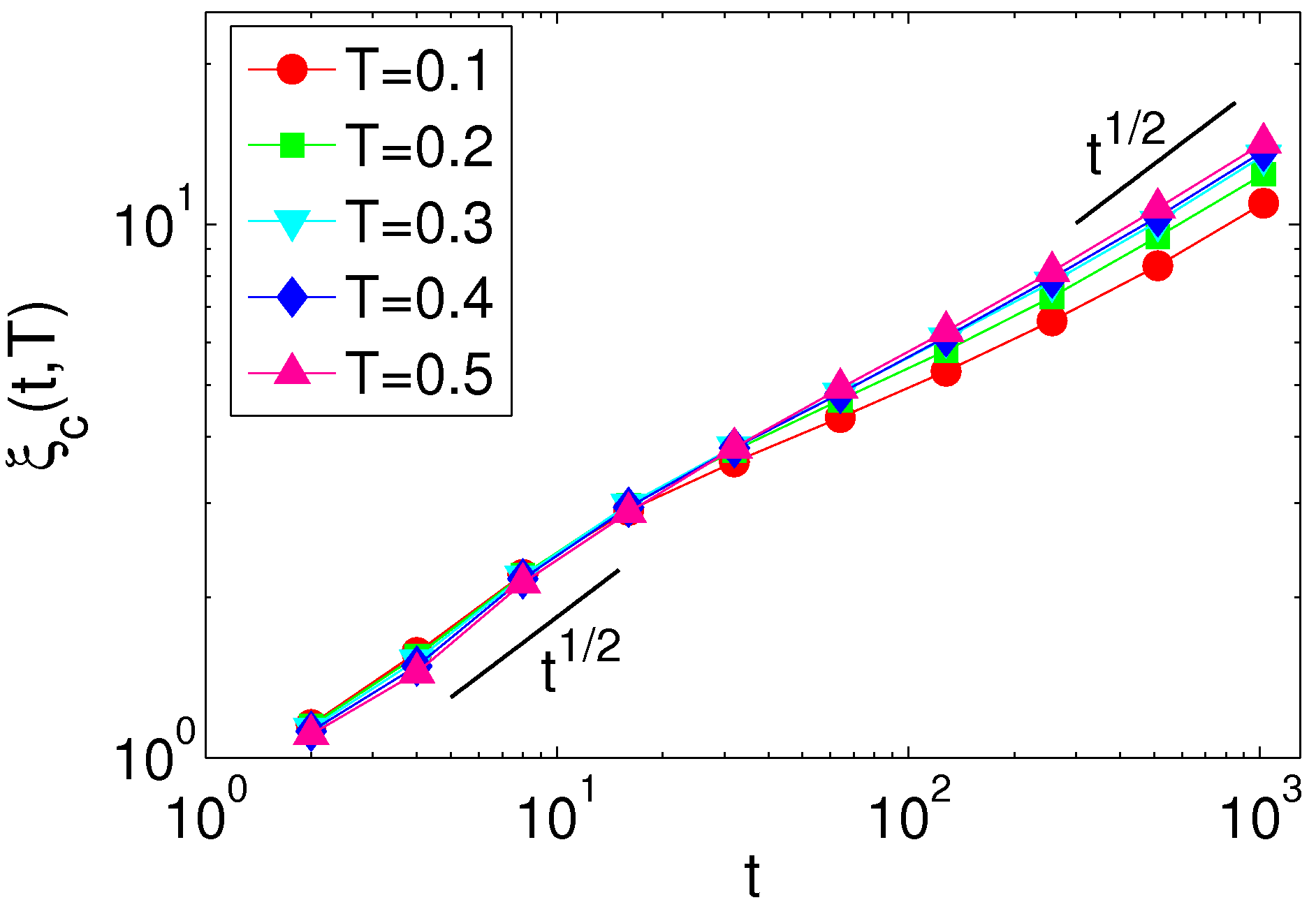}
\end{center}
\caption{(Color online.)  Space-time correlation after an infinitely
  rapid quench from a fully random initial condition into the low
  temperature phase.  (a)~Bare data for $C(r,t)$ at
  different times $t$ after a quench to $T=0.5$ that are given in the
  key.  The inset shows the scaling plot $r^{\eta(T)} C$ against
  $r/\xi_c(t,T)$ using $\xi_c(t,T)$ computed from
$C[\xi_c(t,T),t]=0.3$. (b)~$\xi_c(t,T)$ against $t$ for
  instantaneous quenches to different temperatures. }
\label{fig:growing-length}
\end{figure}

\begin{figure}[h]
\vspace{0.3cm}
\hspace{1.2cm} (a) \hspace{5.5cm} (b)
\begin{center}
\includegraphics[width=0.49\textwidth]{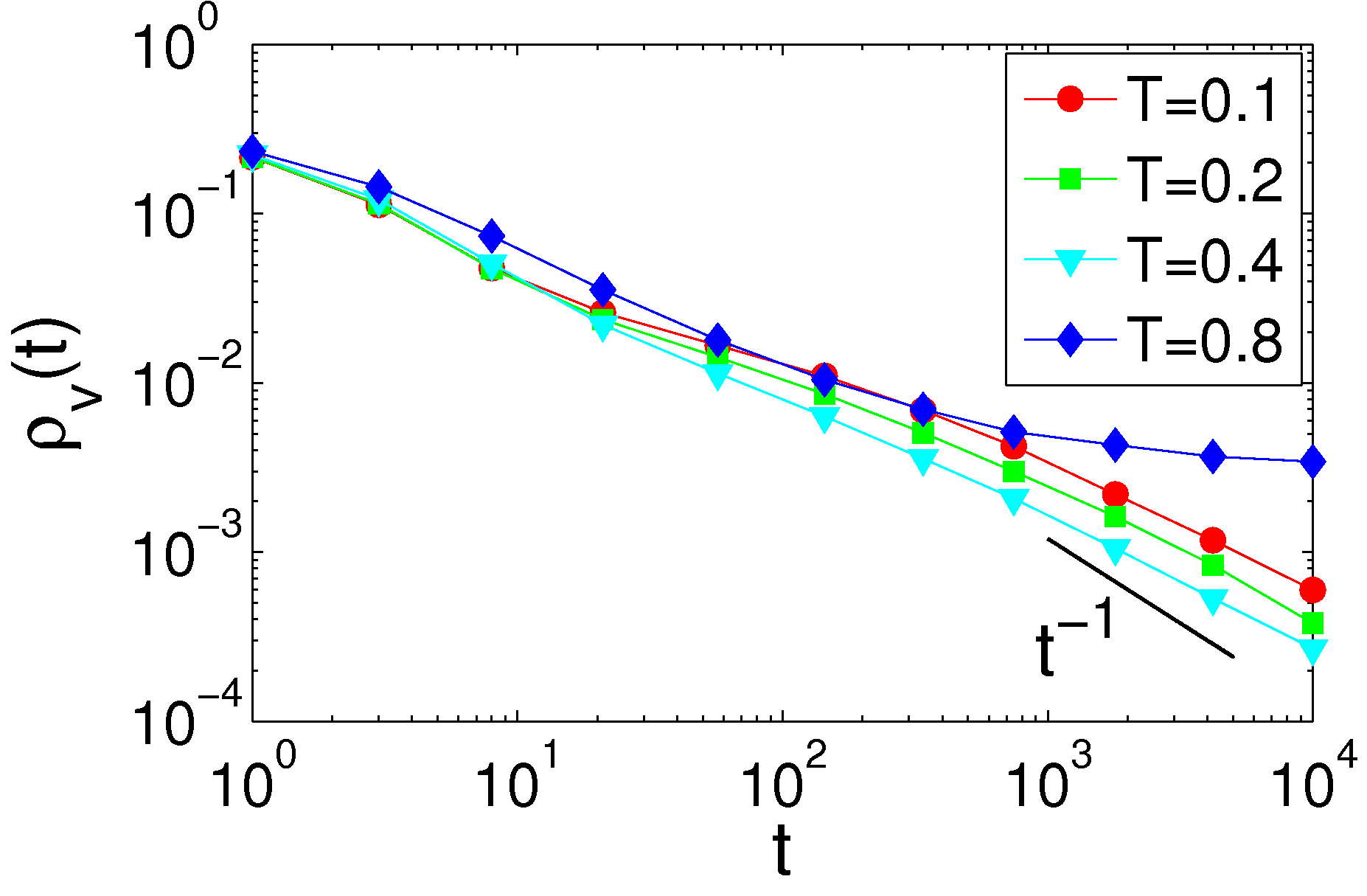}
\includegraphics[width=0.49\textwidth]{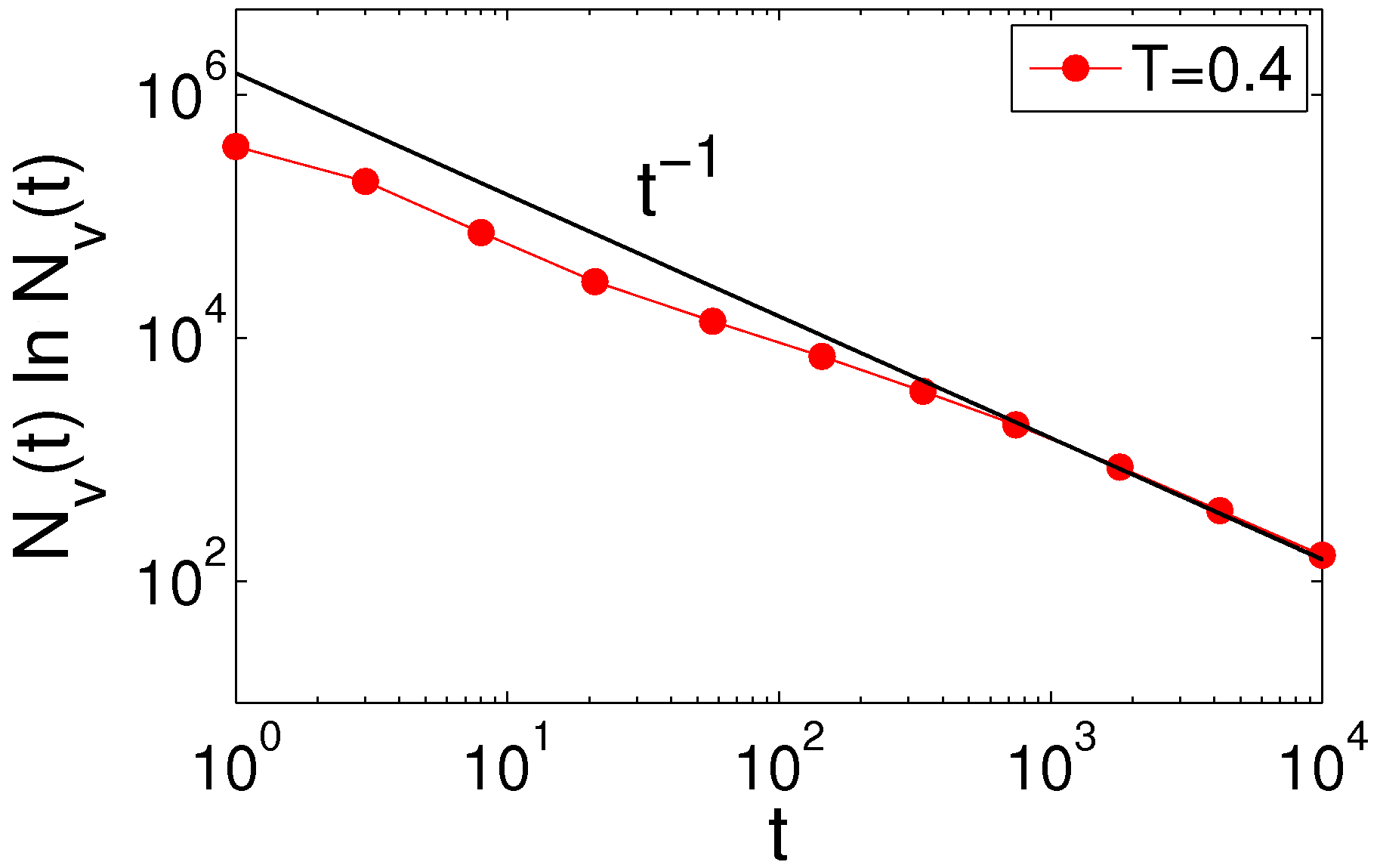}
\end{center}
  \caption{(Color online.) Time-dependence of the total density of
    vortices, $\rho_v(t,T) \equiv N_v(t,T)/L^2$, after an infinitely
    rapid quench from $T_0\to \infty$ to $0\leq T\leq T_{KT}$ given in
    the key. The data are shown in the form $\rho_v(t)$ {\it vs.} $t$
    in (a) and $N_v\ln N_v$ {\it vs.} $t$ at a relatively low
    $T$ in (b).  The former display the deviation from the `difussive' $t^{-1}$ law. 
    The latter confirms the expected logarithmic
    correction to a power law decay. The $T$-dependence and,
    especially, the saturation observed at high $T$ are discussed in
    the text.}
\label{fig:rho-density-quench}
\end{figure}

In Fig.~\ref{fig:growing-length} we extract the 
growing length $\xi(t,T)$ from the decay of the space-time correlation.
In the left panel we show the bare data, and  we 
confirm that $C$ satisfies dynamic scaling by using the value of $\xi_c(t,T)$ 
inferred from the condition 
$C[\xi_c(t,T),t]=0.3$. 
In the right panel we compare the time-dependence of $\xi_c(t,T)$ to
the power law $t^{1/2}$ and we see the expected slower growth at
sufficiently long times.  The small deviations due to the logarithmic
correction will be studied in more detail in
Fig.~\ref{fig:rho-density-quench}.

Next we analyze the growing length as extracted from the density of
vortices.  In the fully random, infinite temperature limit, $\rho_v
\simeq 0.3$.  In equilibrium at $2T_{KT}$ we found a value
slightly smaller than $0.2$ in agreement with data shown
in~\cite{Gupta92}.  After the infinitely rapid quench the very high
density of free vortices present in the initial high temperature
condition decreases. Two processes contribute to the re-organization
of the vortex configuration. On the one hand over-abundant defects
annihilate. On the other hand vortices and anti-vortices bind to
approach the non-vanishing equilibrium density of pairs, $\rho_{eq}
\simeq e^{-2\mu/T}$ with $\mu\simeq 3.77 $~\cite{Gupta92}, that is an
increasing function of $T$.

 According to dynamic scaling, during the out of equilibrium
 relaxation the density of defects in the ordered phase should
 decrease as $\rho_v(t,T) \simeq 1/\xi_v^2(t,T)$ that using
 eq.~(\ref{eq:growth-law}) is equivalent to $\rho_v(t,T) \ln
 \rho_v(t,T) \simeq t^{-1}$~\cite{Huse}.  In
 Fig.~\ref{fig:rho-density-quench} we show our data for all vortices
 in two ways: in the left panel we plot $\rho_v(t)$ as a function of $t$
 for several values of $T$; in the right panel $N_v(t) \ln N_v(t) $ as
 a function of $t$ where $N_v$ is the number of vortices selecting the
 case $T=0.4$. After one MC step the density is very close to the
 initial value $\rho_v(T\to\infty) \simeq 0.3$ but it subsequently
 monotonically decreases in time in all cases.  As already stressed by
 Yurke {\it et al.}~\cite{Huse}, who solved numerically the
 time-dependent Ginzburg-Landau equation with noise, the second
 presentation gives a much better description of data for
 $T\stackrel{<}{\sim} 0.4$, confirming the growth-law
 (\ref{eq:growth-law}).  However, at $T$ close to $T_{KT}$ the number
 of vortices gets close to the equilibrium value, for example
 $\rho_{eq} \simeq 10^{-4}$ at $T=0.8$, and the cross-over to
 equilibrium dynamics is reached within the simulation times. This is
 accompanied by the generation of many short-lived vortex anti-vortex
 pairs, similar to what was observed in~\cite{Huse}, and the data
 deviates from the expected scaling.  The failure of scaling of the
 space-time correlation when $\xi_v$ is used was also mentioned
 in~\cite{Rojas99,Bongsoo}.
 
\begin{figure}[h]
\hspace{1.2cm} (a) \hspace{5.5cm} (b)
\begin{center}
\includegraphics[width=0.48\textwidth]{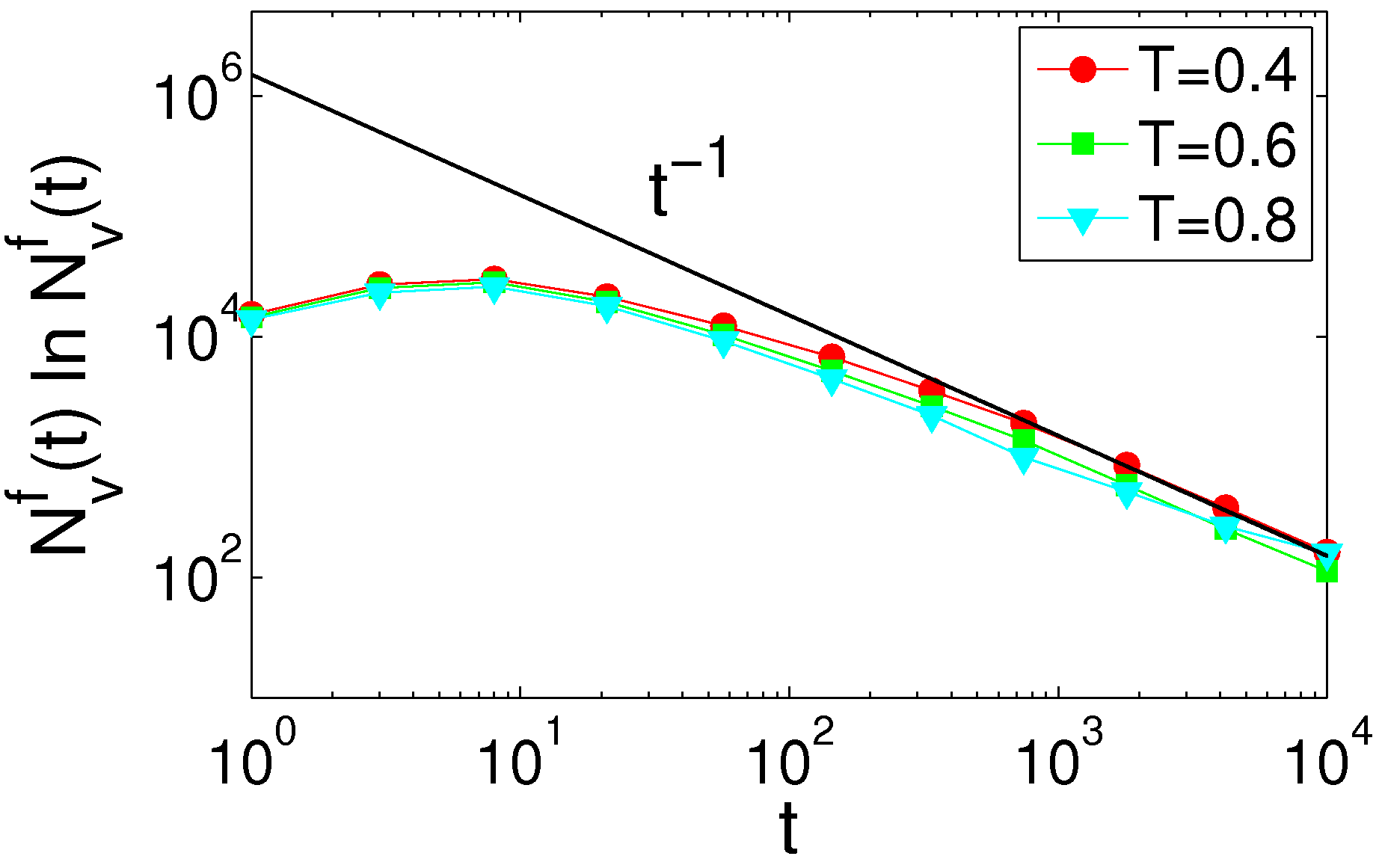}
\includegraphics[width=0.51\textwidth]{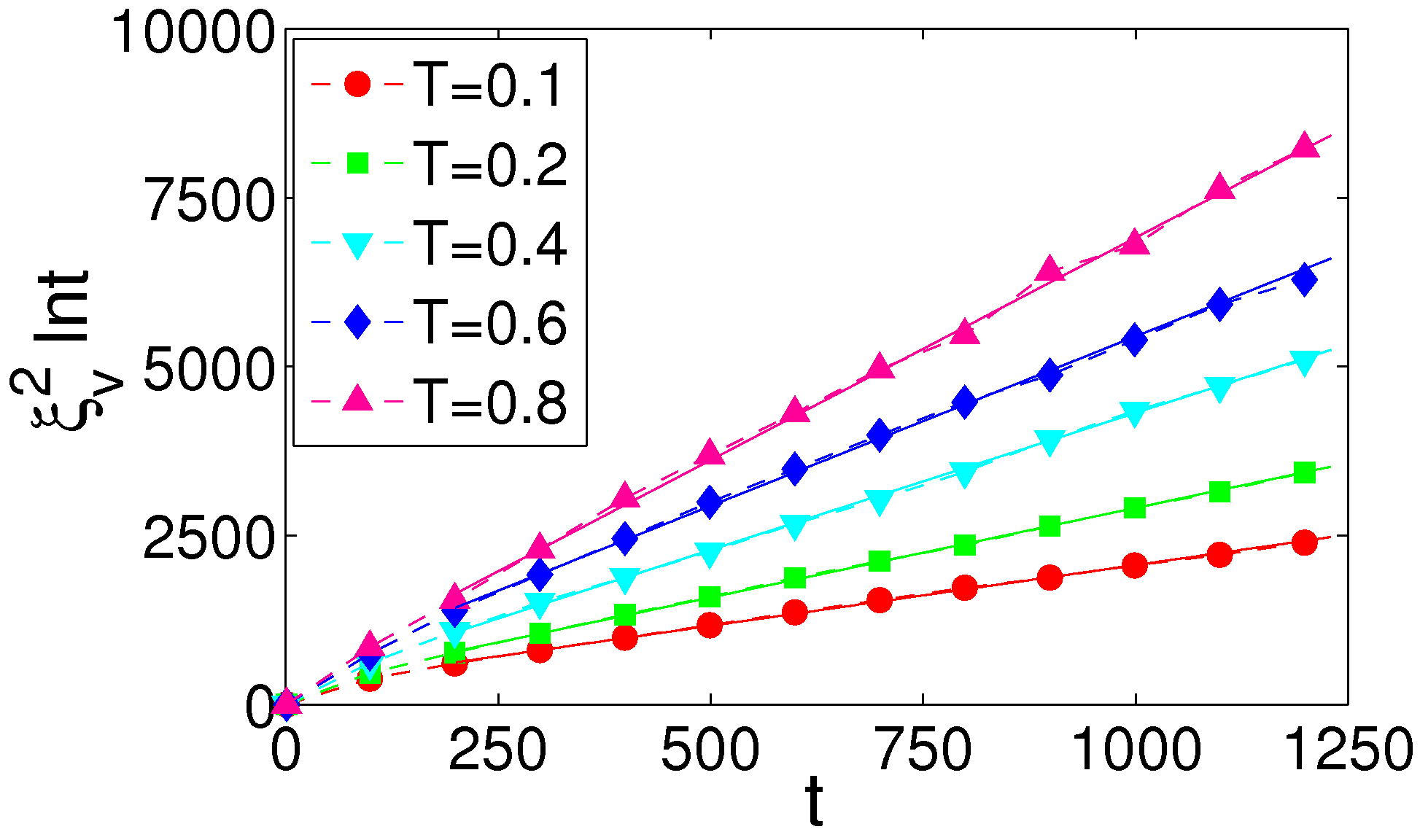}
\end{center}
  \caption{(Color online.) (a) Time-dependence of the density of free
    vortices, expressed in the form $N^f_v(t,T) \ln N^f_v(t,T)$
    against $t$, after an infinitely rapid quench from $T_0\to \infty$
    to $0\leq T\leq T_{KT}$ given in the key. (b) Time-dependence of
    the growing correlation length, defined as $\xi_v \equiv
    (\rho_v^f)^{-1/2}$ for values of $T$ given in the key. Vortices and
    antivortices are considered to be free when the distance to their
    closest neighbour of the opposite vorticity is larger than a
    cut-off that in this figure has been chosen to be $r_c=2$ lattice
    spacings. }
\label{fig:free-density}
\end{figure}

A more precise determination of the growing length at temperatures
close to $T_{KT}$ should be given by the evolution of the density of
free vortices only, $\rho_v^f$. The identification of free vortices
involves, however, some ambiguity. We used a simple-minded form that
consists in counting as free all vortices and antivortices that do not
have a neighbour of opposite vorticity at Euclidean distance $ r \leq
r_c$. By varying the parameter $r_c$ from $1$ to $2$ we found that the
second choice gives very good results, shown in
Fig.~\ref{fig:free-density} in the form $N_v^f \ln N_v^f$ against $t$
for $T=0.4, \ 0.6, \ 0.8$. (In~\cite{Wysin} $r_c$ is chosen to be equal 
to $\xi(t,T)$.) Consistently, $\xi_v^f \simeq \xi_v$ at
sufficiently low temperatures (say, $T\leq0.4$) but the two depart at
higher temperatures, with the former yielding the correct out of
equilibrium correlation length and being the one that ensures dynamic 
scaling. 

\subsubsection{Temperature dependence}

\begin{figure}[h]
\vspace{0.3cm}
\hspace{1.3cm} (a) \hspace{5.5cm} (b) 
\begin{center}
\includegraphics[width=0.54\textwidth]{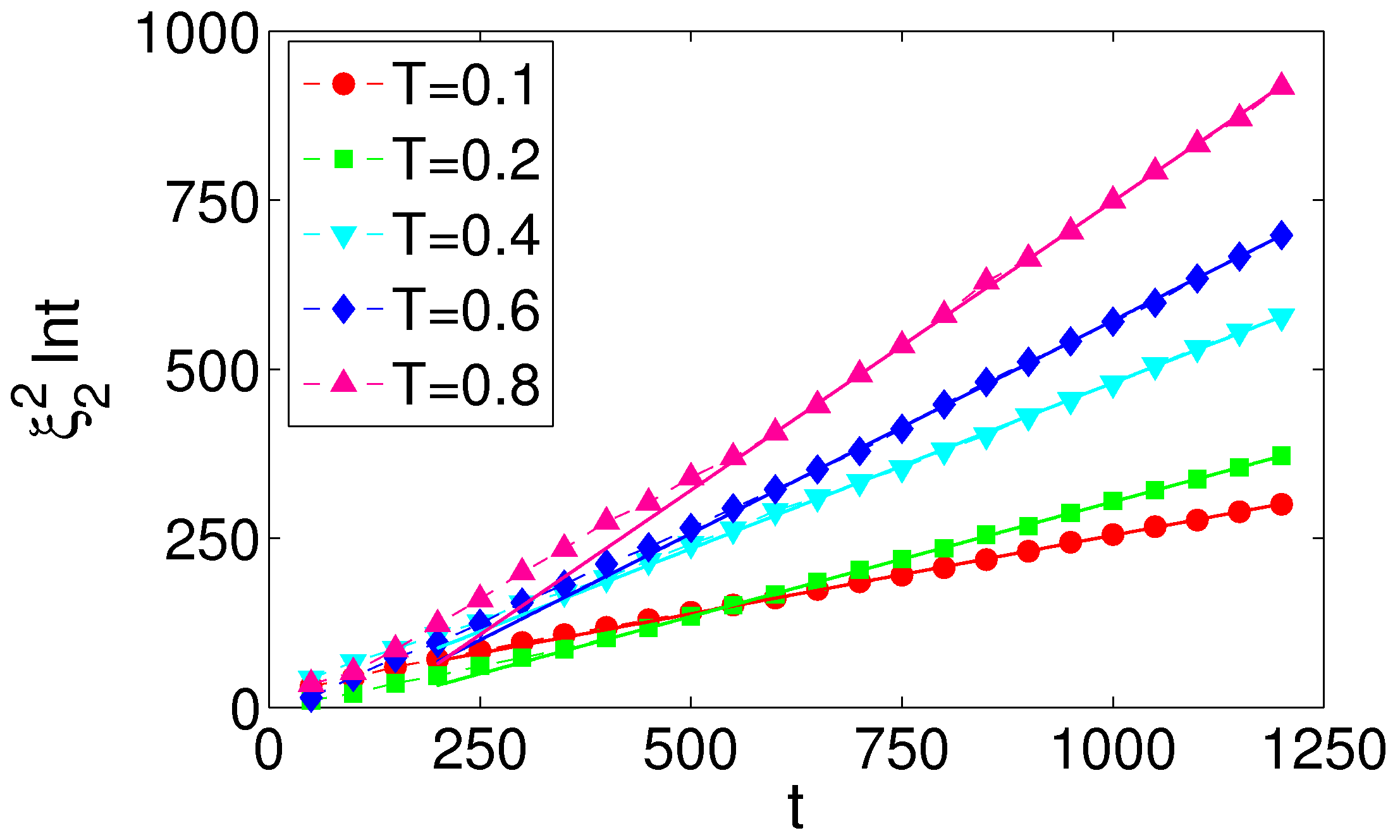}
\includegraphics[width=0.45\textwidth]{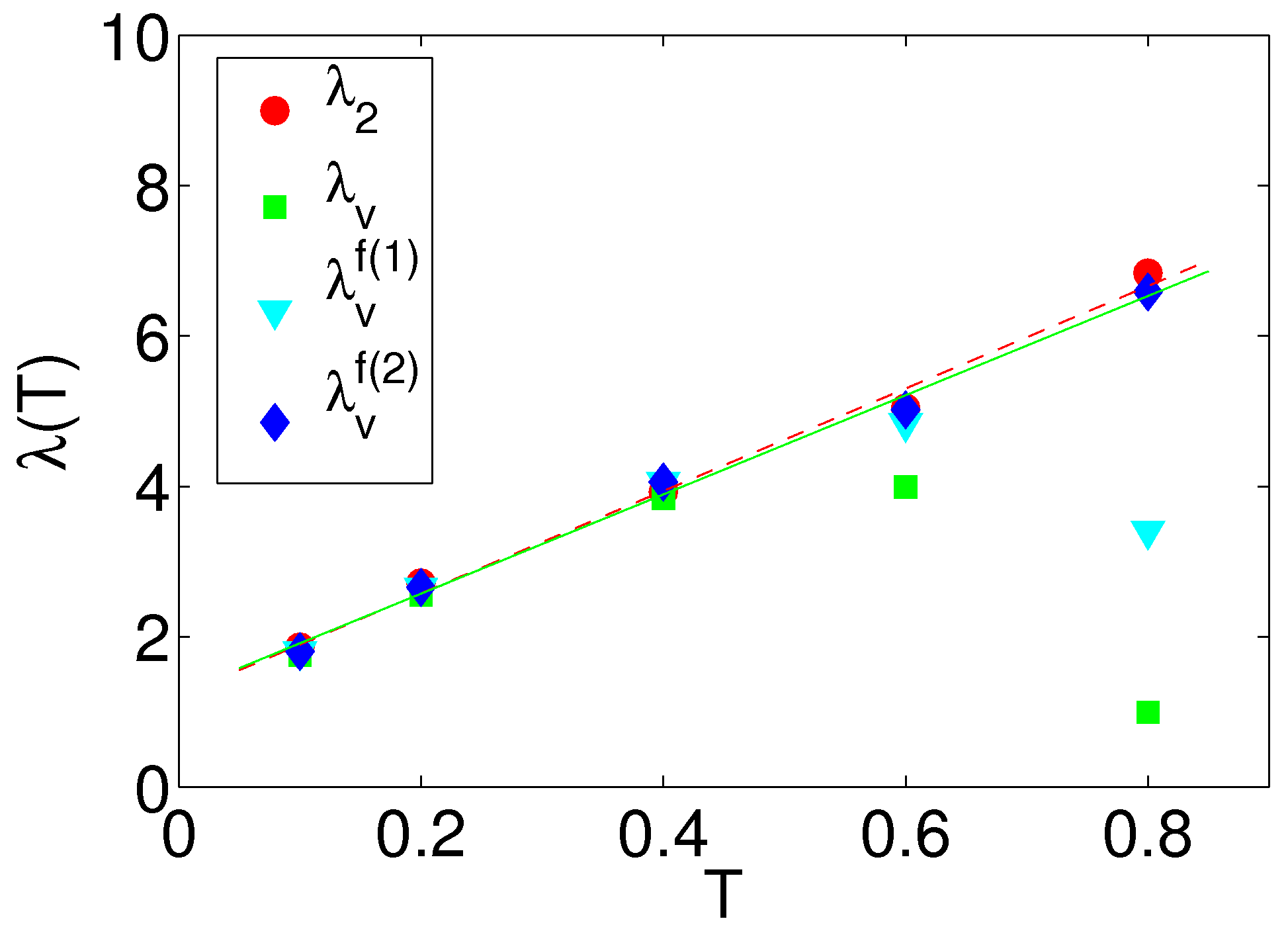}
\end{center}
  \caption{(Color online.) Temperature dependence of the
    parameter $\lambda(T)$ in the growth law (\ref{eq:growth-law})
    after an instantaneous quench from infinite temperature into the
    ordered phase.  (a) The growing length determined as
    $\xi_{2}(t,T)$. The slope in the long-time limit is $\lambda(T)$
    and it increases with increasing $T$ for all $T$'s studied. The
    time-scale $t_0(T)$ cannot be determined due to the long
    transient. (b) $\lambda(T)$ using four determinations: 
    $\xi_2$, the total density of vortices and the density of
    free vortices with  $r_c=1$ and $r_c=2$. The agreement between the 
first and the last one is very good.}
    \label{fig:T-dep}
\end{figure}

The $T$-dependence in $\lambda$ and $t_0$ in Eq.~(\ref{eq:growth-law})
has not been fully established yet.  The assumption that at finite
temperature the dynamics on scales that are shorter than $\xi(t,T)$
should be described, in the limit of large $\xi(t,T)$, by renormalized
spin-wave theory~\cite{Bray94,Bray00} suggests $\lambda(T)=\rho_s(T)
\Gamma(T)$ and $t_0(T)=a^2/[\rho_s(T) \Gamma(T)]$ at all $T\leq
T_{KT}$.  $a$ is the lattice spacing or microscopic cut-off, $\rho_s$
the renormalized spin-wave stiffness and $\Gamma$ the renormalized
kinetic coefficient in the time-dependent Ginzburg-Landau model.  The
numerical data in~\cite{Luo98} analyzed with an effective power-law
$\xi(t,T) \simeq \overline\lambda(T) t^{1/z(T)}$ yield an effective
$T$-dependent exponent that slowly increases with decreasing $T$.
This is consistent with a microscopic time-scale $t_0(T)$ in
Eq.~(\ref{eq:growth-law}) that also weakly increases upon decreasing
temperature. This in turn implies that $\lambda(T)$ is a weakly
increasing function of $T$.  We analyzed the $T$-dependence of the
parameters $\lambda$ and $t_0$ by plotting
\begin{equation}
\xi^2(t,T)\ln t \qquad \mbox{against} \qquad t
\end{equation}
in Fig.~\ref{fig:free-density}~(b) and~\ref{fig:T-dep}~(a), 
using $\xi^f_{v}$ and $\xi_{2}$, respectively. 
Since
the microscopic time $t_0$ cannot be determined precisely due to the
relatively long transient needed to reach the asymptotic law, and the
fact that it yields a sub-dominant contribution anyhow, we fix it to
one and determine the remaining free-parameter $\lambda$. The data fall
on straight lines with slope $\lambda(T)$. For all temperatures
simulated, $\lambda(T)$ is an increasing function of temperature. As
the temperature approaches $T_{KT}$, $\lambda(T)$ as obtained from the
density of all vortices decreases, see Fig.~\ref{fig:T-dep}. As
already explained, this is an artifact of not having distinguished
free from bound vortices. The agreement between the trend obtained
from $\xi_2$ and $\xi_v^f$ is very satisfactory. The dashed line 
is a linear fit to the datapoints, 
$\lambda(T) \simeq \lambda_0 + c T$ with $\lambda_0 \simeq 1.2$ and $c=6.8$. 

\begin{figure}[h]
\hspace{0.8cm} (a) \hspace{5.5cm} (b)
\begin{center}
\includegraphics[width=0.49\textwidth]{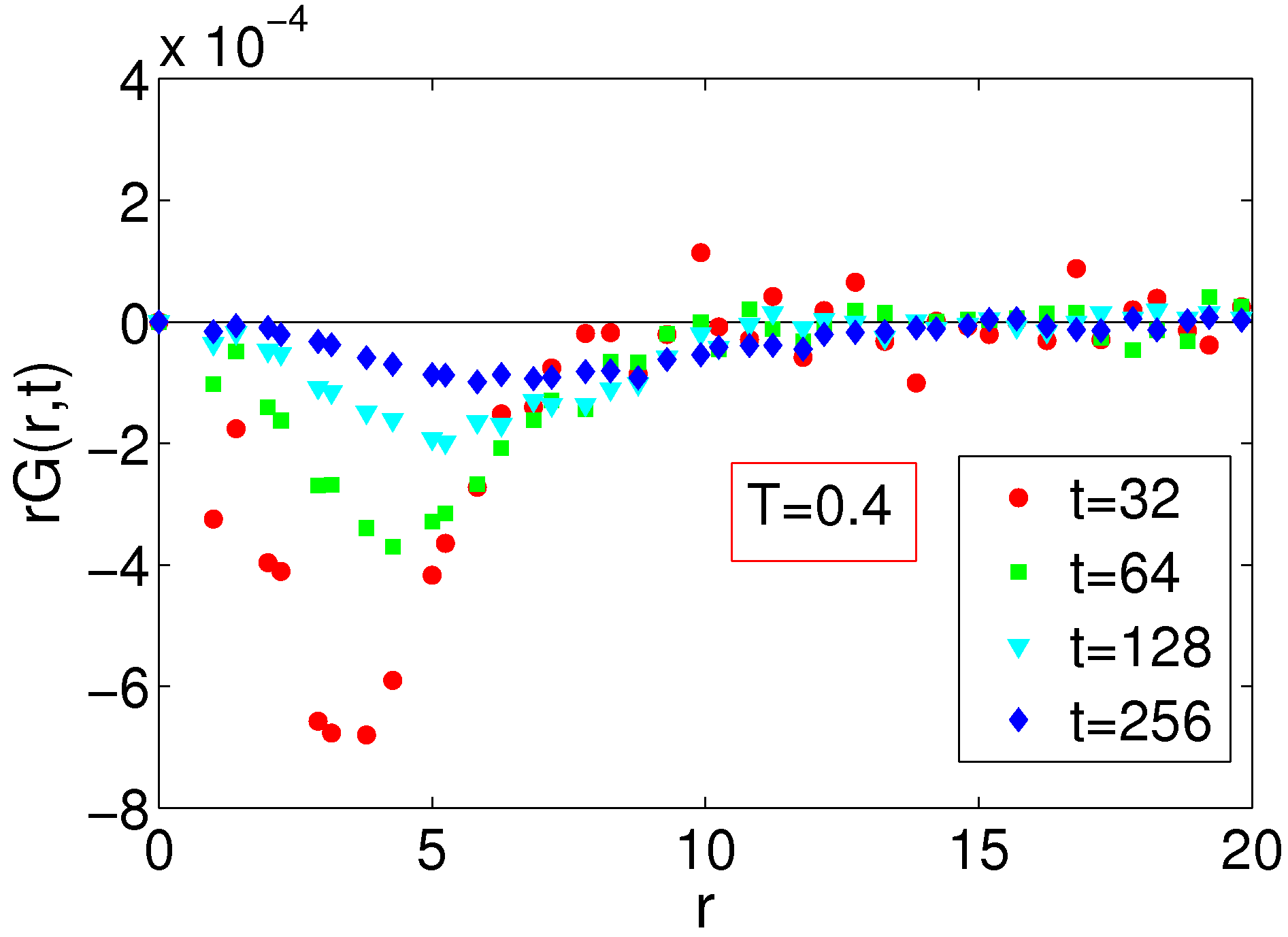}
\includegraphics[width=0.49\textwidth]{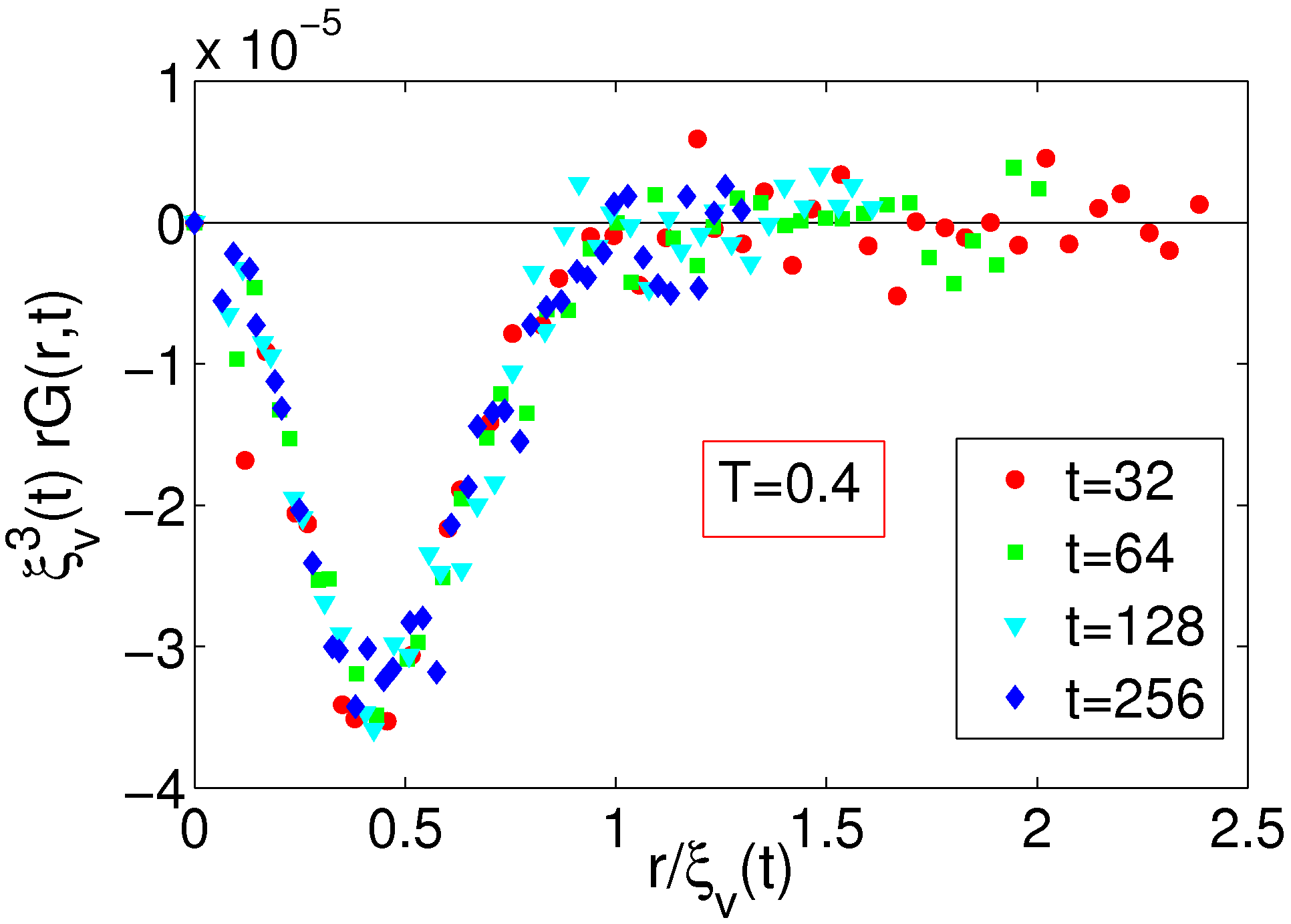}
\end{center}
  \caption{(Color online.) Polarity correlation function.  (a) $r
    G(r,t)$ against $r$ at different times $t$ after the quench into
    the low-temperature phase. (b) Scaling plot $\xi_v^3(t) rG(r,t)$
    against $r/\xi_v(t)$ with $\xi_v(t)$ the length extracted from the
    analysis of the density of free vortices. The numerical data have
    been smoothened.}
    \label{fig:Gr}
\end{figure}

The $T$-dependence of $t_0$ and $\lambda$ are consistent with 
Berthier {\it et al}'s recognition that the Monte Carlo generated 
vortex configurations at a given waiting-time after a quench to
different temperatures are rather different.  At high $T$ (although
below $T_{KT}$) the vortices tend to form pairs, as they should in
equilibrium below $T_{KT}$, while at lower $T$ there are more free
vortices. The suggestion is that the dynamics are faster at higher $T$
(since vortex diffusion is faster) than at lower $T$~\cite{Berthier}.

\subsection{Spatial structure}

Further information on the vortex anti-vortex pair structure is
obtained from the analysis of pair correlation functions that ignore
or take into account the polarity of the vortices.  In the former case
we define a variable $n_i$ that equals $1$ if the site $i$ is occupied
by a defect irrespective of its polarity and $0$ otherwise. In the
latter the variable $n_i$ takes three values; $n_i=1$ if the site is
occupied by a vortex, $n_i=-1$ if it is occupied by an anti-vortex,
and $n_i=0$ if there is no defect. The correlation function is then
defined as $G(r,t)= \langle n_i(t) n_j(t)\rangle$ where $r=|\vec
r_i-\vec r_j|$ in both cases. The first type of correlation does not
yield additional information on the behaviour of the system with
respect to $C(r,t)$ studied above. Instead, the latter has a minimum
at the preferred distance between vortices and
anti-vortices~\cite{Liu}.  In Fig.~\ref{fig:Gr} we study $G(r,t)$ at
several instants after the quench, showing in (a) the raw data in the
form $r G(r,t)$ against $r$ and in (b) the scaling plot
$\xi^{3}_v(t,T) rG(r,t)$ against $r/\xi_v(t,T)$ with $\xi_v(t,T)$ the
correlation length obtained from the analysis of the density of free
vortices (similar results are obtained using $\xi_2$).  The data
confirm that this quantity is well described by dynamic scaling. The
scaling function has a minimum at $r\simeq 0.48 \ \xi_v$ and it
oscillates with very small amplitude around zero at long distances, a
feature observed experimentally~\cite{Israel} but not captured by the
approximate theory in~\cite{Liu}.

\subsection{Pair distribution function}

\begin{figure}[h]
\hspace{0.5cm} (a) \hspace{5.5cm} (b)
\\
\begin{center}
\includegraphics[width=0.45\textwidth]{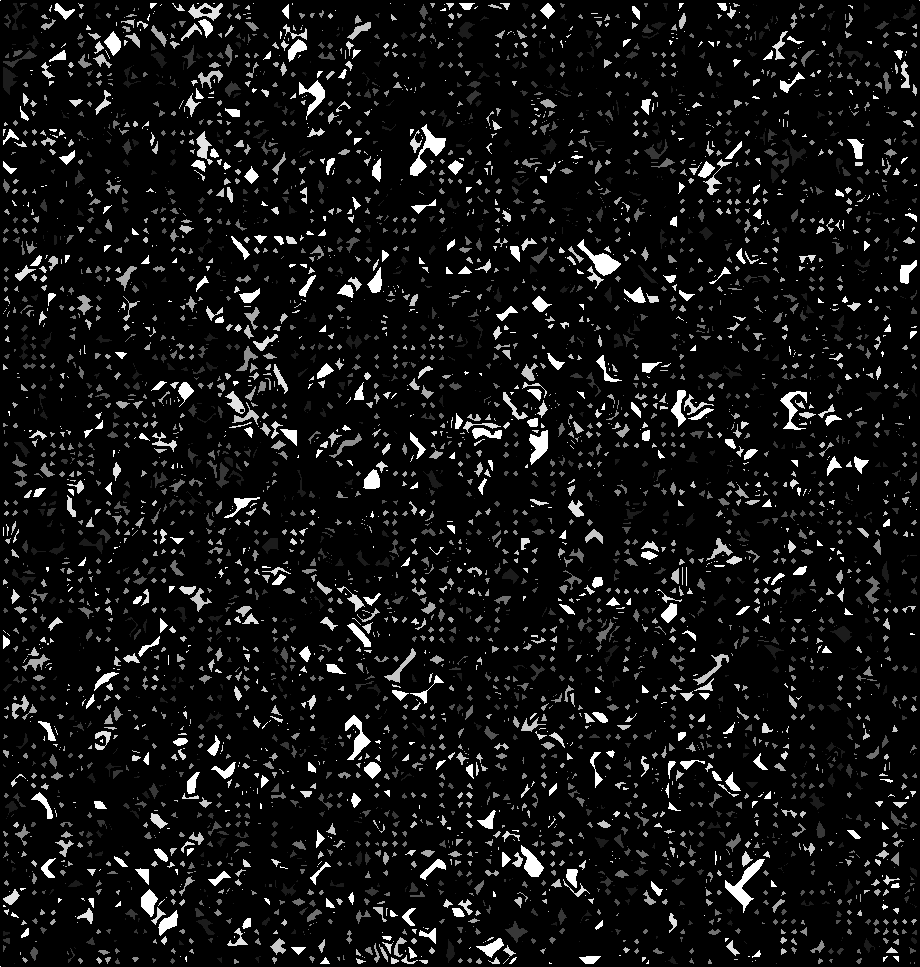}
\hspace{0.1cm}
\includegraphics[width=0.45\textwidth]{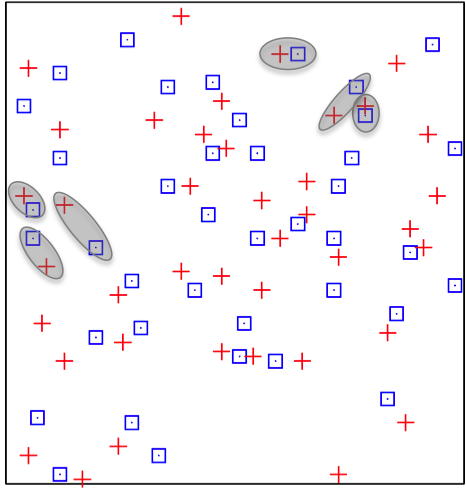}
\end{center}
  \caption{(Color online.) Defect configuration at $t=100$ MCs after
    quenching the sample to $T=0.4$ in the ordered phase. The
    configurations are shown as Schlieren patterns over the full
    lattice ($L=100$) on the left panel.  The gray scale is proportional to
    $\sin^2(2\phi)$ with $\phi$ the angle formed by the local spin and
    the $x$ axis.  On the right panel, the vortex configurations at
    time $t=100$ are shown. Vortices are represented by (red) pluses and
    antivortices by (blue) squares.  The total number of vortices in the
    system is $N_{v}(t=100)=80$. We graphically show a few pairing of
    vortices and antivortices. }
    \label{fig:Schileren-quench}
\end{figure}

 MC simulations give us access to the distances between vortices and
 anti-vortices and, in principle, to their pairing.  The assignment of
 pairing is, however, a hard problem not free from ambiguity. We used
 the following simple algorithm. Given a configuration, we first
 computed and ordered all distances between vortices and
 anti-vortices.  However, the lattice structure implies that, for
 sufficiently high density, some vortex (or anti-vortex) could be at
 equal distance from two (or more) anti-vortices (or vortices). In
 these cases we chose the pairing at random and we continued the
 procedure with the remaining defects. With this method we are not
 sure of finding the optimal pairing but, statistically, we expect
 all these pairings to be equivalent as the distribution of
 distances is concerned, a feature that we verified numerically.
 An example of the pair-distance construction is shown in 
 Fig.~\ref{fig:Schileren-quench}. In panel (a) the configuration at $t=100$ MCs
 after a quench from $T_0\to\infty$ to $T=0.4$ is 
 represented as a Schlieren pattern in which a gray scale is
proportional to $\sin^2(2\phi)$ with $\phi$ the angle formed by the
local spin and the $x$ axis. Each vortex emanates eight brushes
of alternate black and white color. In panel (b) the vortices and anti-vortices 
are shown with (red) pluses and (blue) squares. A few pairings, as obtained 
with our simple algorithm,  are highlighted 
in grey. 

\begin{figure}[h]
\hspace{0.8cm} (a) \hspace{5.5cm} (b)
\begin{center}
\includegraphics[width=0.49\textwidth]{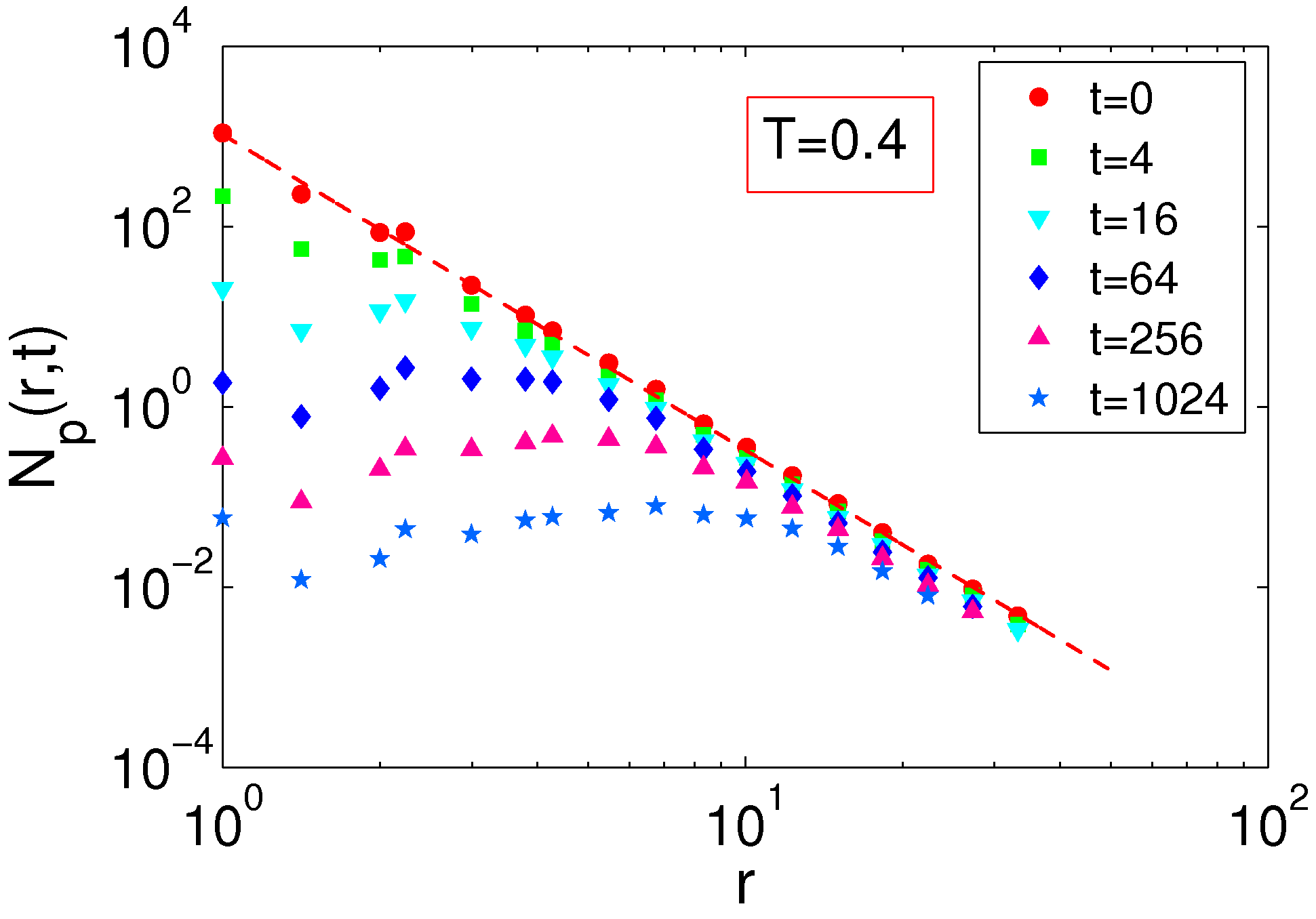}
\includegraphics[width=0.49\textwidth]{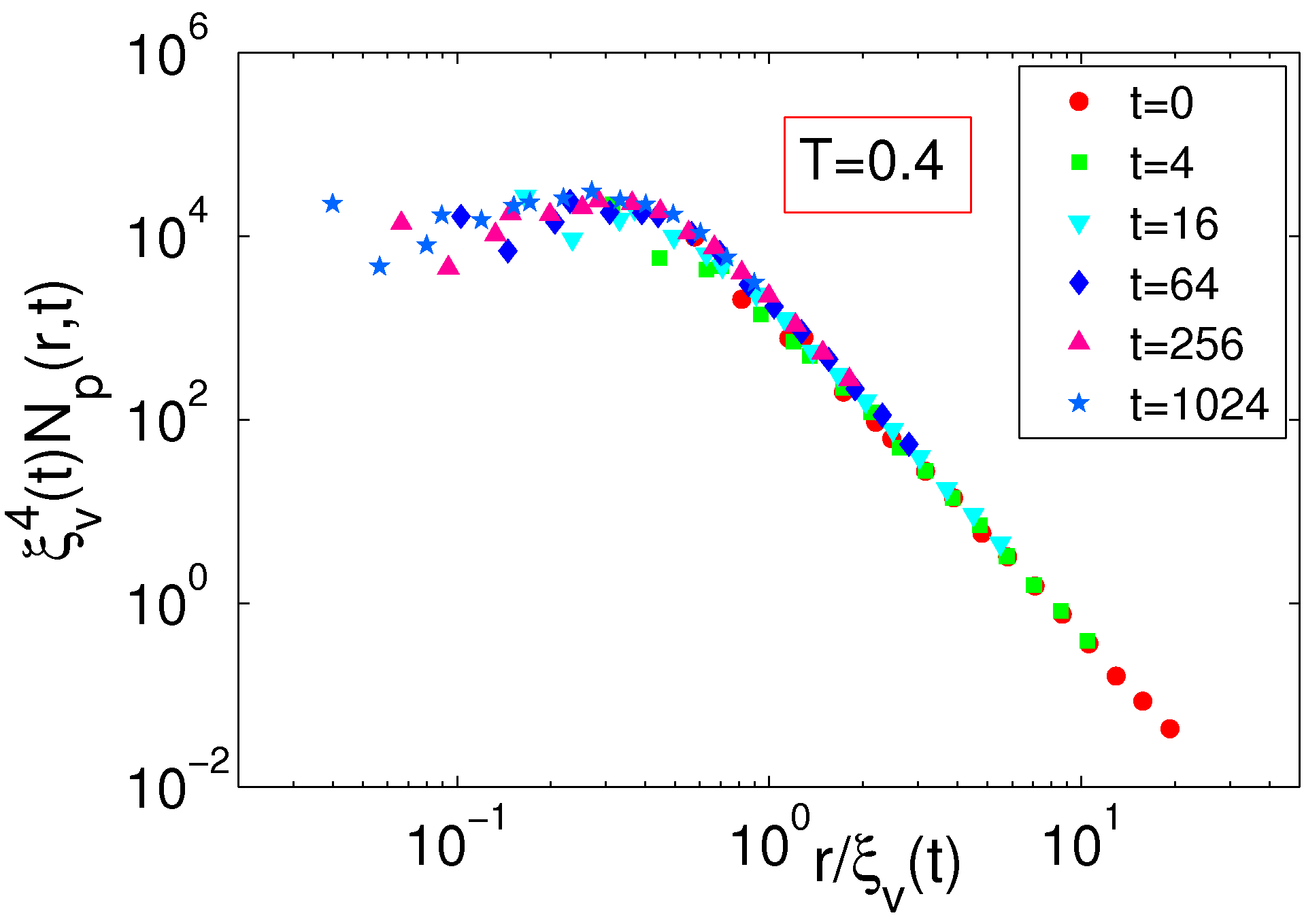}
\end{center}
  \caption{(Color online.) Number of vortex-antivortex pairs, $N_p$,
    as defined in the text, after the quench from $T\to \infty$ to
    $T=0.4$.  (a) $N_p$ as a function of their distance $r$ at different
    times given in the key. The dashed line is a guide-to-the-eye
    representing the power $r^{-3.5}$. (b) Scaling plot.}
 \label{fig:pairs-distribution}
\end{figure}

Figure~\ref{fig:pairs-distribution} displays the evolution of the
distribution of pair distances after a quench from infinite
temperature to  $T=0.4$.  The log-log plot in
panel~(a) demonstrates that the initial  probability distribution 
function (pdf) is a power law, $\simeq
r^{-\tau}$ with $\tau\simeq 3.5$.  Similarly to what was found in the study of geometric
properties of coarsening in the Ising universality
class~\cite{Alberto}, small linear scales -- as compared to $\xi_v$ --
change significantly in time while larger scales do not.
This means that the tail of the distribution remains the initial  power-law for
$r\gg \xi_v(t)$ while the weight on smaller scales decreases
significantly as a function of time. Panel (b) presents the 
scaling plot, $\xi_v^{4} N_p(r,t)$ against $r/\xi_v(t)$, 
in which one confirms that dynamic scaling is 
well satisfied. The crossover at $r\simeq \xi_v(t)$ is clear in the figure.
The kind of oscillating features at very short distances, $r\stackrel{<}{\sim}
3$, might be due to the discrete lattice. 

The data in Fig.~\ref{fig:pairs-distribution} is to be compared to
results shown in Fig.~1 of Ref.~\cite{Chu}. The latter were obtained
using the Fokker-Planck equation for the density of pairs of
separation between $r$ and $r+dr$ proposed in~\cite{Ambegaokar}. This
equation is tailored to hold at $T\leq T_{KT}$, it does not take
into account the effect of free vortices, and it does not capture
the logarithmic correction to the diffusive growing length. The quench 
is done from
equilibrium at $T_{KT}$, where the distribution is the power law
$r^{-\tau}$ with $\tau \simeq 4.70$, to a very low $T$. The
qualitative features of the two sets of data are the same. However,
the quantitative properties are not.  The initial conditions, and
hence the initial distributions, are different. The time-dependent
part has a peak in our case and it is totally flat in~\cite{Chu}.
Our pdf satisfies scaling with respect to the correct growing length.

\section{Dynamics after a slow cooling}
\label{sec:annealing}

\begin{figure}[h]
\begin{center}
\includegraphics[width=0.49\textwidth]{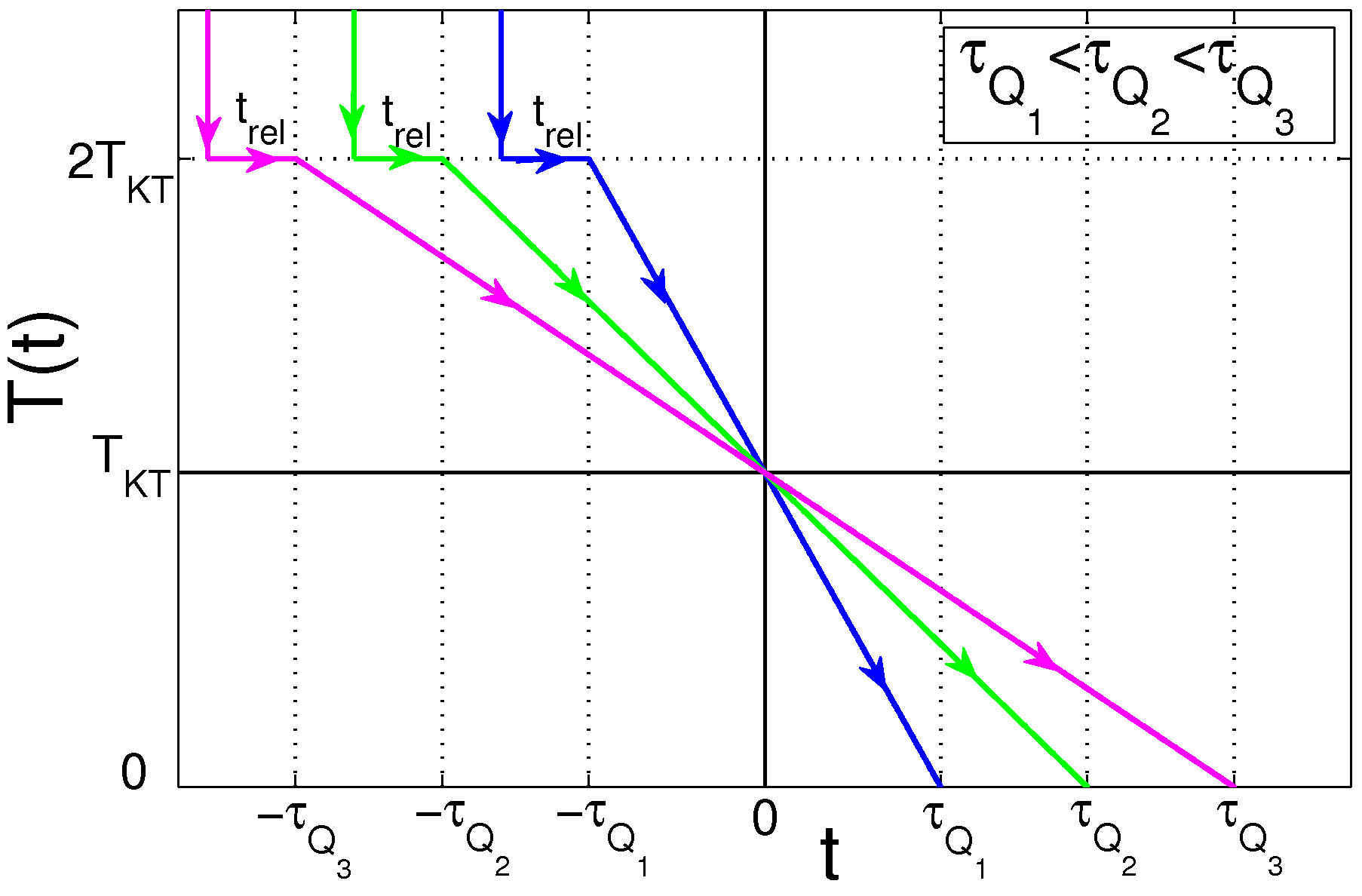}
\end{center}
\caption{(Color online.) Sketch of the cooling procedures used.}
\label{fig:sketch}
\end{figure}

In this Section we discuss the evolution of the system after a slow
cooling from high $T$ to its low $T$ phase. The cooling procedure used
is graphically represented in Fig.~\ref{fig:sketch} and it consists in
the following.  We suddenly quench the sample from $T\to\infty$
(random initial condition) to $T_0$. In practice, we chose
$T_0=2T_{KT}$.  We let the system evolve at this temperature a large
number of MCs, typically $t_{rel}=2\times 10^{5}$ to $6\times 10^{5}$ 
depending on the system size. Equilibration
occurs relatively rapidly far from $T_{KT}$ so for all purposes we can
assume the system to be in equilibrium at $2T_{KT}$. We then cooled
the sample by changing the temperature linearly in time according to
\begin{equation}
T(t) = T_{KT} \left(1- t/\tau_Q \right)  \;
\;\;\; -\tau_Q \leq t \leq \tau_Q
\; .
\label{eq:linear-cooling}
\end{equation}
Within this time reference, annealing from $2T_{KT}$ 
starts at $t=-\tau_Q$.  At negative times $t$ the system is
above the phase transition, at $t=0$ it reaches $T_{KT}$ and at
positive times it enters the ordered phase. $\tau_Q$ is the inverse
cooling rate. At $t=\tau_Q$ the environmental temperature vanishes and
times are naturally bounded by $\tau_Q$. See Fig.~\ref{fig:sketch} for
a graphical representation of these cooling rate procedures.
Although more
complicated cooling procedures are also of interest, and have been 
analyzed in detail in the $1d$ Glauber Ising chain~\cite{Krapivsky}, we 
shall not discuss them here. 

\subsection{Snapshots}

A first intuitive understanding of the defect density left over in the ordered
phase after a slow annealing through the phase transition is obtained
by looking at the spin configurations. 

\begin{figure}
\hspace{1cm} (a) \hspace{5.5cm} (b)
\begin{center}
\includegraphics[width=0.45\textwidth]{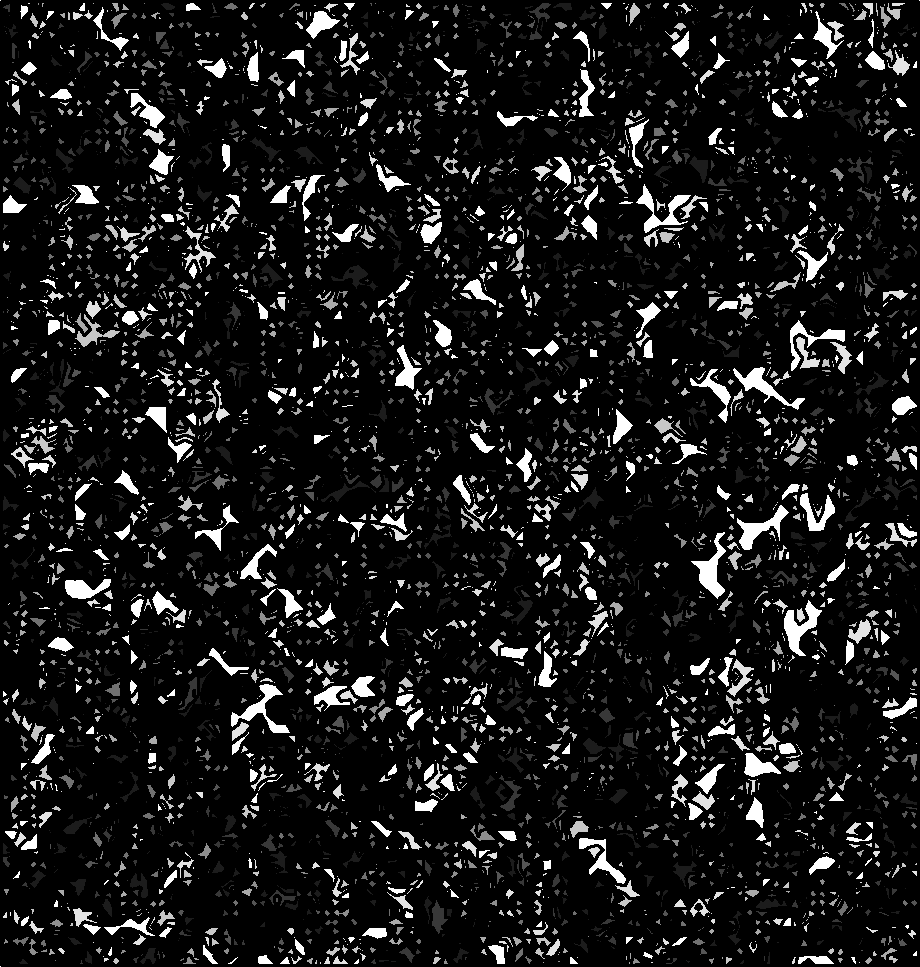}
\hspace{0.1cm}
\includegraphics[width=0.45\textwidth]{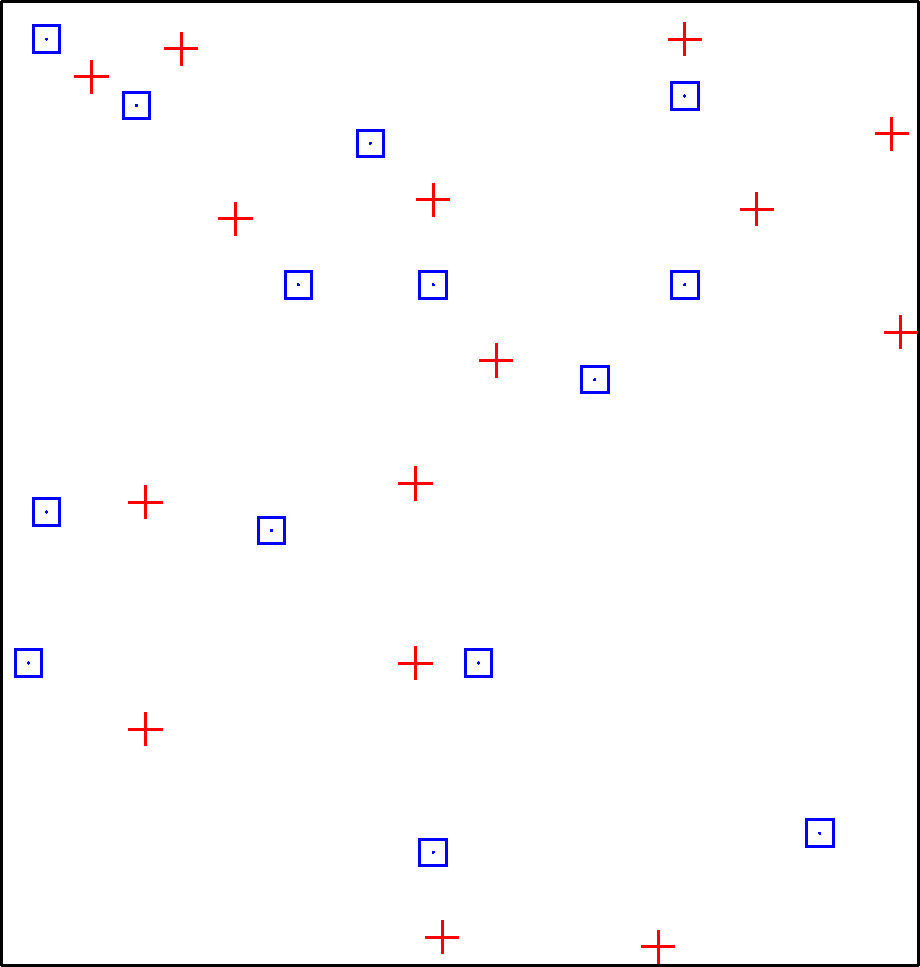}
\end{center}
\hspace{1cm} (c) \hspace{5.5cm} (d)
\begin{center}
\includegraphics[width=0.45\textwidth]{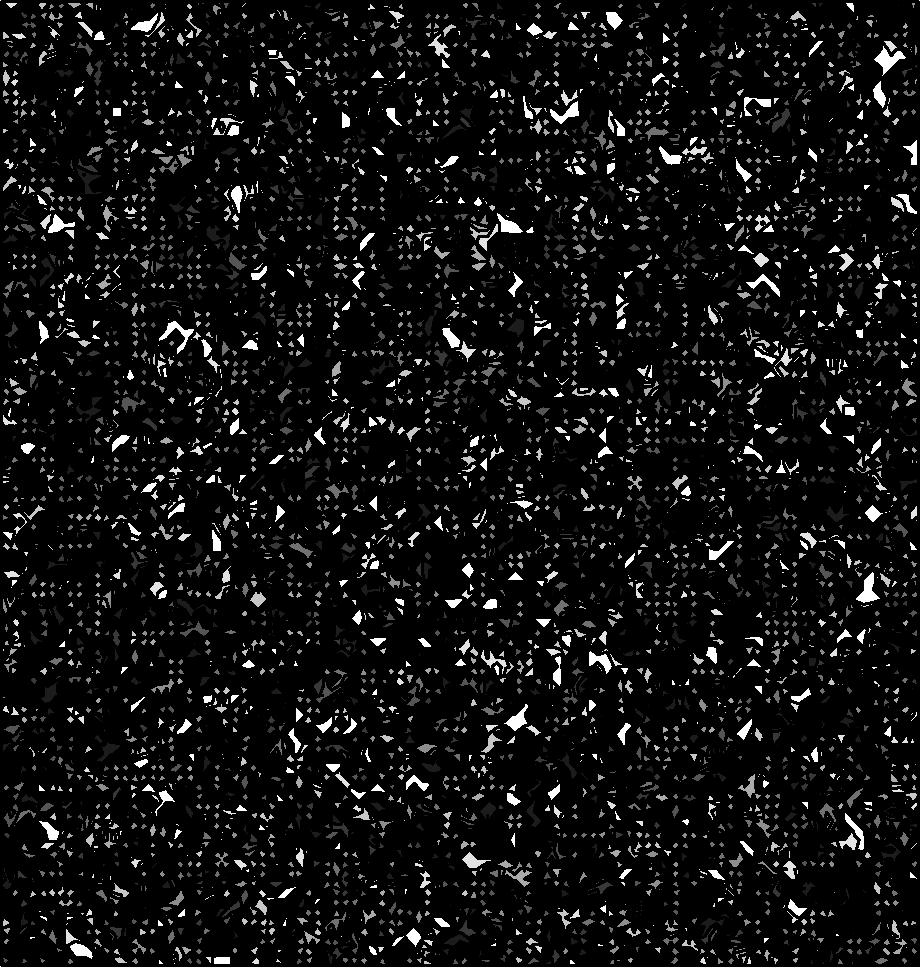}
\hspace{0.1cm}
\includegraphics[width=0.45\textwidth]{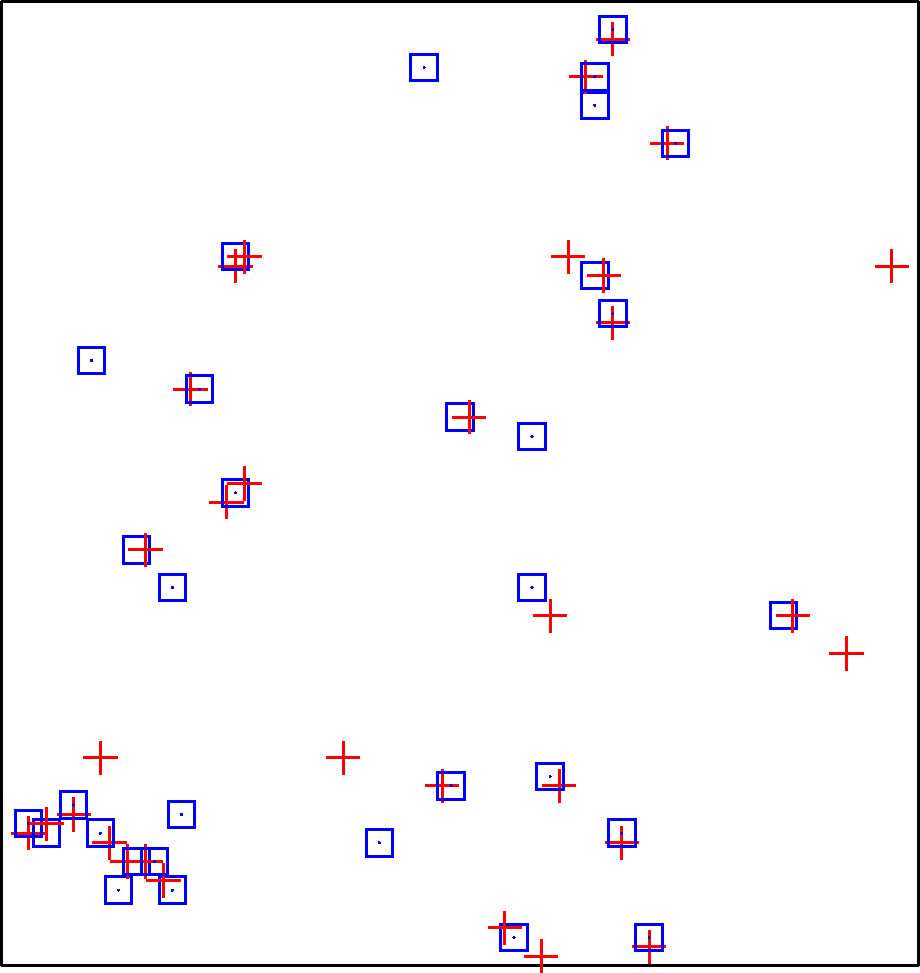}
\vspace{0.2cm}
\end{center}
  \caption{(Color online.) Comparison between the defect configuration
    at $t=200$MCs$<\min(\tau_Q)$ after annealing the sample into the
    ordered phase with two different inverse cooling rates $\tau_Q=256$ MCs
    and $2048$ MCs. We show parts of linear size $100$ of the full
    lattice $L=256$. The configurations are shown as Schlieren
    patterns on the left [(a) and (c)].  The gray scale is
    proportional to $\sin^2(2\phi)$ with $\phi$ the angle formed by
    the local spin and the $x$ axis.  On the right side, the vortex
    configurations at time $t=200$ are shown [(b) and (d)]. Vortices
    are represented as (red) pluses and antivortices by (blue)
    squares.  The total numbers of vortices in the system are
    $N_{v}=186$ for the fast cooling in which $T=0.2$ is reached and
    $N_v=352$ for the slow one in which the configuration is measured
    at $T=0.8$. }
\label{fig:Schlieren1}
\end{figure}

In Fig.~\ref{fig:Schlieren1} we
display the system configuration on a section of linear size $L=100$ 
of a system with size $L=256$ at $t=200$ MCs after having
crossed the phase transition with two inverse cooling rates
$\tau_Q=256, \ 2048$ MCs.  
Note that the temperature at which the snapshots are taken is 
different, $T(t=200, \tau_Q=256)=0.2$ [(a) and (b)] and 
$T(t=200, \tau_Q=2048)=0.8$ [(c) and (d)]. 
The figure demonstrates that the 
density of vortices increases with increasing $\tau_Q$, with $N_v=186$ 
in the first case and $N_v=352$ in the second. 

\begin{figure}
\hspace{1cm} (a) \hspace{5.5cm} (b)
\begin{center}
\includegraphics[width=0.45\textwidth]{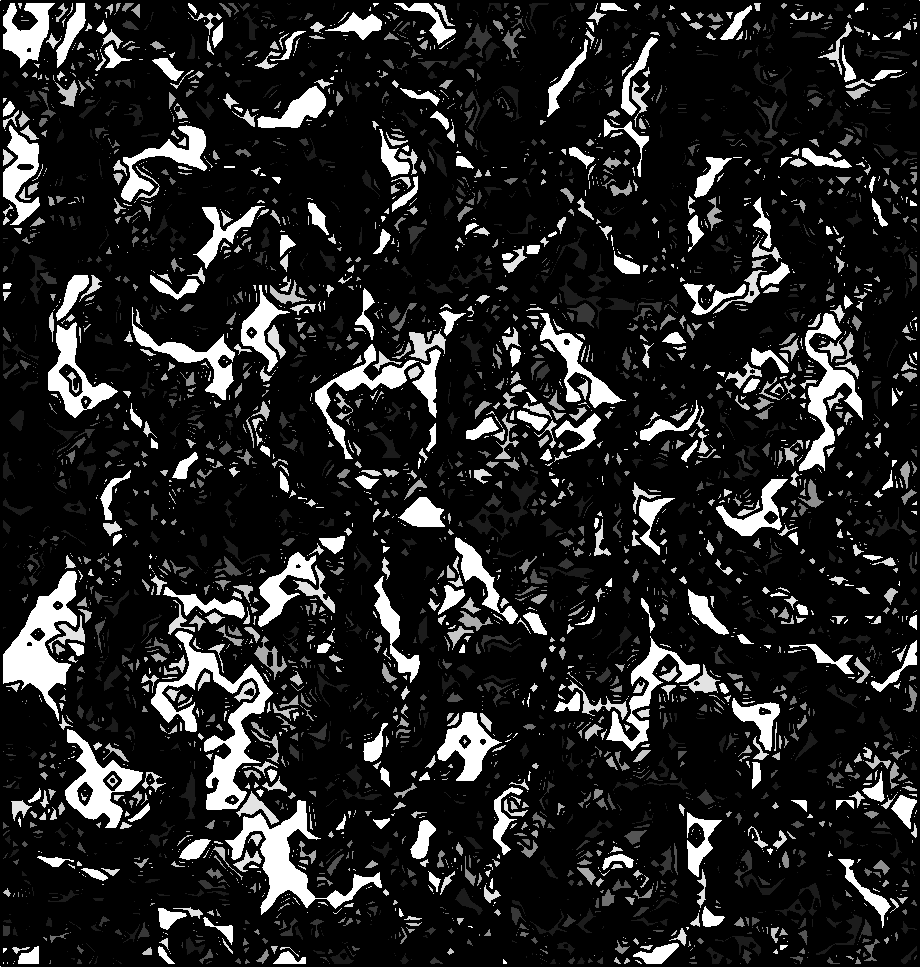}
\hspace{0.1cm}
\includegraphics[width=0.45\textwidth]{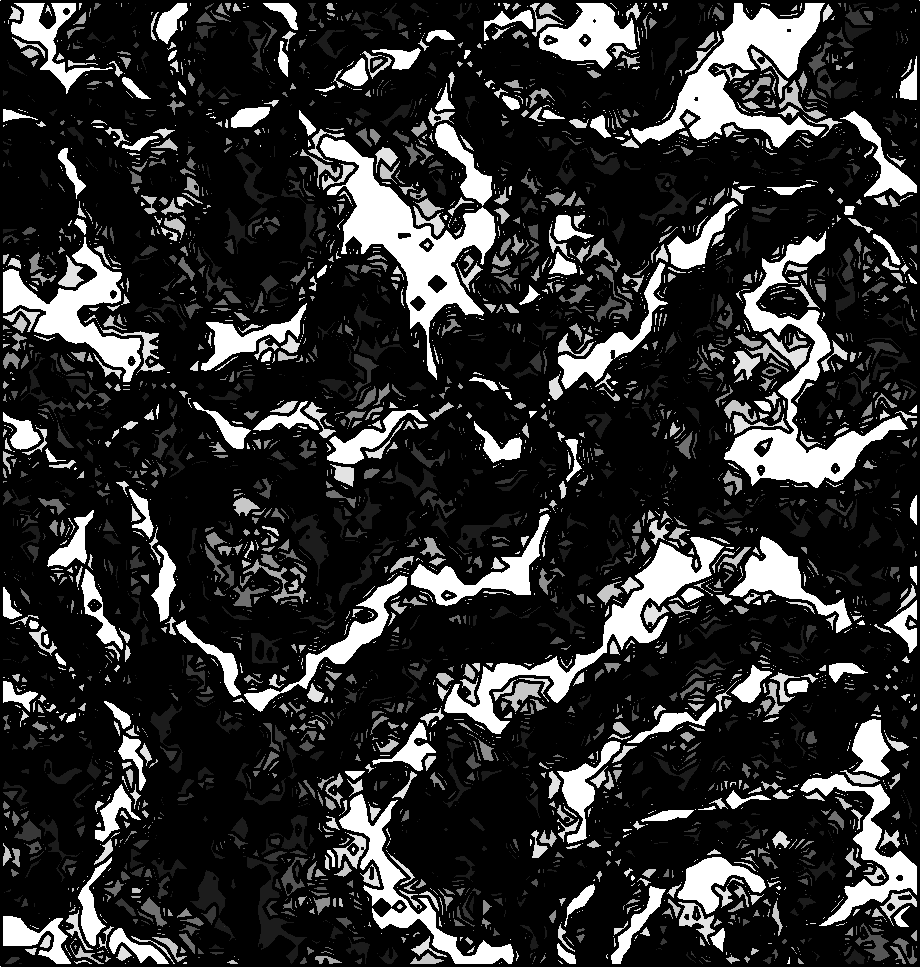}
\end{center}
\hspace{1cm} (c) \hspace{5.5cm} (d)
\begin{center}
\includegraphics[width=0.45\textwidth]{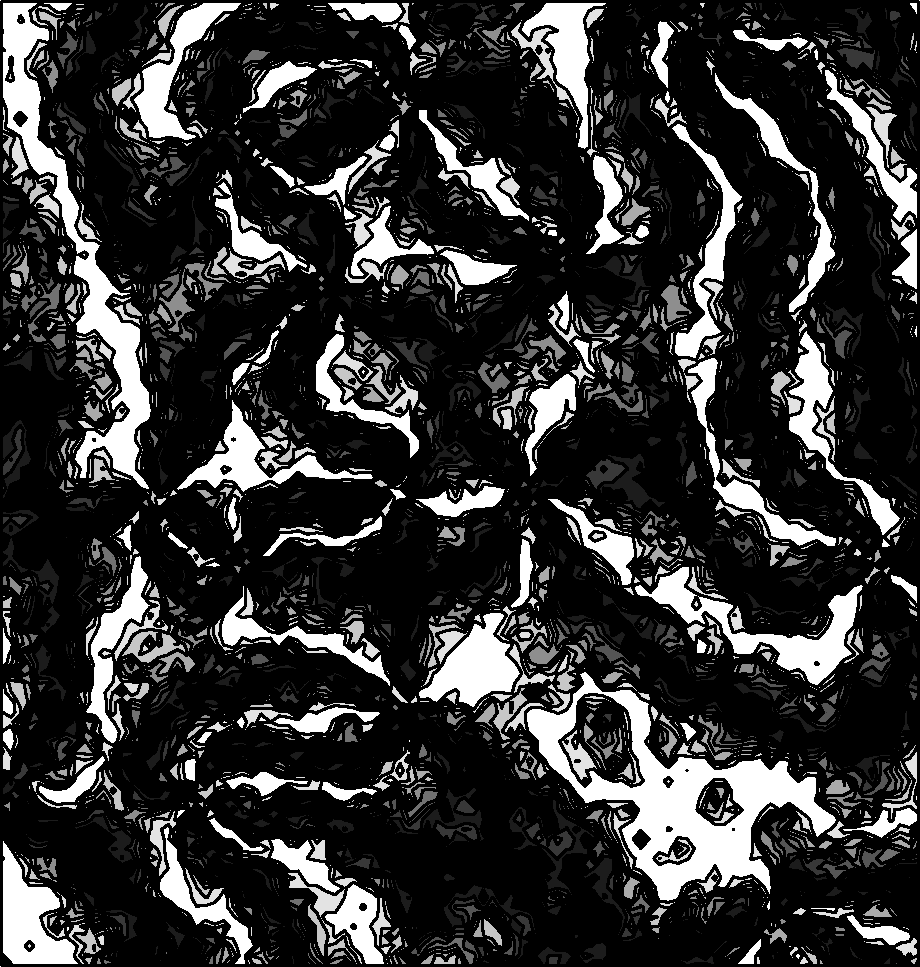}
\hspace{0.1cm}
\includegraphics[width=0.45\textwidth]{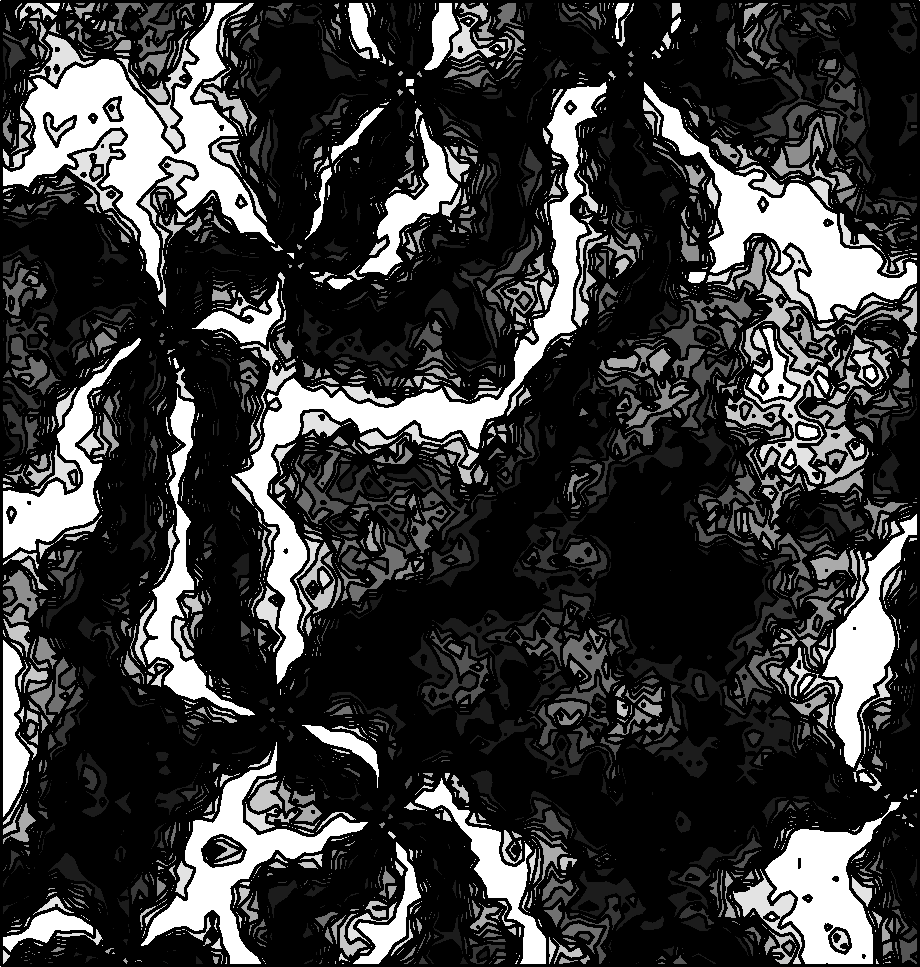}
\vspace{0.2cm}
\end{center}
  \caption{(Color online.) Comparison between the defect configuration
    at $t=\tau_Q$ after annealing the sample to $T=0$ with four
    different inverse cooling rates $\tau_Q=256$ (a), $512$ (b), $1024$ (c)
    and $2048$ (d).  We show a section of linear size $100$ of
    the full lattice with $L=256$.  The total number of defects
    decreases with increasing $\tau_Q$ and are $N_v=180, \ 92, \ 68,
    \ 42$ from (a) to (d).}
\label{fig:Schlieren2}
\end{figure}

In Fig.~\ref{fig:Schlieren2} we show four configurations at
$t=\tau_Q$, when the temperature has reached the value zero, with
$\tau_Q=128, \ 512, \ 1024, \ 2048$ MCs from (a) to (d),
respectively. It is clear from the figure that the density of vortices
decreases with increasing $\tau_Q$. We measured $N_v=180, \ 92,
\ 68, \ 42$ for increasing $\tau_Q$.  In the following we shall make
these statements quantitative.

\subsection{Equilibrium -- out of equilibrium crossover}

The annealing procedure occurs by the following steps. In the high temperature 
phase the system is plagued with free vortices and anti-vortices. The
initial state thus contains a large number of free defects. 
Let us assume that the system is in equilibrium at $T_0$ ($=2T_{KT} $ in our simulation).
Under the cooling procedure, if this is slow enough, it   
evolves adiabatically, {\it i.e.}, it
remains instantaneously in equilibrium at $T(t)$. The equilibrium
correlation length is swept in time according to
\begin{equation}
\xi_{eq}(t) \simeq a_\xi e^{b_\xi |t/\tau_Q|^{-\nu}}
\; 
\end{equation}
and the density of vortices, estimated from $\rho_v(t) \simeq \xi^{-2}_{eq}(t)$, 
decreases exponentially:
\begin{equation}
\rho_v(t) \simeq a_\xi^{-2} \ e^{-2 b_\xi | t/\tau_Q|^{-\nu}}
\; . 
\end{equation}
However, this regime cannot last for ever, it crosses over to a different, much 
slower one, when the system falls out of equilibrium due to the fact that the relaxation time 
becomes much longer than the parameter variation imposed by the cooling rate procedure. 
Indeed, using $\tau_{eq}$ as derived in~(\ref{taueq-Tdep}) and replacing 
$T(t)$ by its time-dependence, one finds
\begin{eqnarray}
\tau_{eq}(t) \simeq a_\tau e^{z b_{\xi} |t/\tau_Q|^{-\nu}} |t/\tau_Q|^{-\nu}
\; .
\end{eqnarray}
The
system can no longer follow the pace of evolution set by the changing
conditions.  The crossover is estimated as the moment when the
relaxation time $\tau_{eq}$ reaches the characteristic time of
variation of the temperature\footnote{In the $1d$IM with Glauber dynamics, a model
  with an exponential divergence of the relaxation time close to the
  critical temperature $T_c=0$, Krapivsky estimates $-\hat t$ for
  generic cooling rates by comparing the time needed to reach the
  critical point to the relaxation time~\cite{Krapivsky}. For a linear cooling the two 
definitions are identical.} $\Delta T/d_t \Delta T = -\hat
t$~\cite{Zurek}.  The crossover occurs above the critical
temperaure, $\Delta T\equiv T-T_{KT}>0$ and $\hat t>0$.
The negative  time $-\hat t$ is computed
by assuming that the crossover occurs close to the transition, so that
the critical scaling (\ref{eq:equil-corr-length}) determines the
equilibrium relaxation time $\tau_{eq}$.   One then has 
$|-\hat t| \simeq \tau_{eq}(-\hat t)$ and
\begin{equation}
\hat t \simeq a_\tau \ (\hat t/\tau_Q)^{-\nu} \ 
e^{
 z {b_\xi \  (\hat t/\tau_Q)^{-\nu}
}}
\; .
\label{eq:cross-over}
\end{equation}
The $\tau_Q$ dependence of $\hat t$ can be obtained numerically through the Lambert 
W function, {\it i.e.}, the inverse of $f(x) = xe^x$.
An analytic argument to get the leading $\tau_Q$ dependence relies on
recasting~(\ref{eq:cross-over}) as 
\begin{equation}
(1+\nu) \ln x + \ln(\tau_Q/a_\tau) \simeq b_\xi z x^{-\nu}
\qquad\mbox{with}\qquad x\equiv \hat t/\tau_Q
\; . 
\label{eq:cross-over-x}
\end{equation}
Assuming that the solution is such that $x\to 0$ with $\tau_Q/a_\tau\to \infty$ 
we find, in first approximation,
\begin{equation}
x\equiv \frac{\hat t}{\tau_Q} =\frac{\hat T-T_{KT}}{T_{KT} }
\simeq 
\left(\frac{b_\xi z}{\ln ( \tau_Q/a_\tau)} \right)^{1/\nu}
\; , 
\label{eq:first-approx-x}
\end{equation}
that is consistent with the assumption. $\hat t$ is smaller than
$\tau_Q$ -- due to the logarithmic correction -- and this is in
agreement with the fact that the system should fall out of equilibrium
after the beginning of the annealing procedure at $t=-\tau_Q$.
Plugging now the `first order' solution~(\ref{eq:first-approx-x}) in
Eq.~(\ref{eq:cross-over-x}), we find a further logarithmic correction:
\begin{equation}
x\equiv \frac{\hat t}{\tau_Q} =\frac{\hat T-T_{KT}}{T_{KT} }
\simeq 
\left\{\frac{b_\xi z}{\ln \left[ 
\frac{\tau_Q}{a_\tau} \left( \frac{b_\xi z}{\ln (\tau_Q/a_\tau)}\right)^{(1+\nu)/\nu} \right]} \right\}^{1/\nu}
\; . 
\label{eq:log-corr}
\end{equation}
We compared the numerical solution to Eq.~(\ref{eq:cross-over}) to the
estimate (\ref{eq:log-corr}) for different $\tau_{Q}$. For small
$a_{\tau}$ and $b_{\xi}\sim1$ the error is less than $10\%$ for
$\tau_{Q}\stackrel{>}{\sim}200$.  Taking into acount the full form
(\ref{eq:log-corr}) the correlation length at the crossover reads
\begin{equation}\label{eq:xi-hat}
\hat \xi \simeq a_\xi \left\{ \frac{\tau_Q}{a_\tau} \ 
\left[ \frac{1}{\ln (\tau_Q/a_\tau)} \right]^{(1+\nu)/\nu}\right\}^{1/z}
\; . 
\end{equation}
The length $\xi(t(T), T)$ coincides with the equilibrium correlation
length $\xi_{eq}$ for $T(t)>\hat T$, when $\xi(\hat t, \hat T)$ reaches
$\hat \xi$, and $\xi(t(T),T) < \hat \xi$ for $T(t)<\hat T$. The
crossover length, $\hat \xi$, increases and the crossover temperature,
$\hat T$, decreases with increasing $\tau_Q$. Consequently, one can
get closer to the critical point evolving in equilibrium, and thus
reach longer equilibrium correlation lengths, for longer $\tau_Q$s.

 For the sake of comparison, recall that in a conventional
 second-order phase transition with power-law divergencies, these
 quantities are both power laws~\cite{Zurek}: $\hat t/\tau_Q =(\hat
 T-T_{KT})/T_{KT} \simeq (\tau_Q/a_\tau)^{-1/(1+\nu z_{eq})} $ and
 $\hat \xi \simeq a_\xi (\tau_Q/a_\tau)^{\nu/(1+\nu z_{eq})}$. $\hat
 t$ is also smaller than $\tau_Q$ in this case.

\subsection{Numerical estimate of the crossover length }
\label{sec:crossover}

\begin{figure}[h]
\vspace{0.2cm}
\hspace{1cm} (a) \hspace{5.5cm} (b)
\begin{center}
\includegraphics[width=0.49\textwidth]{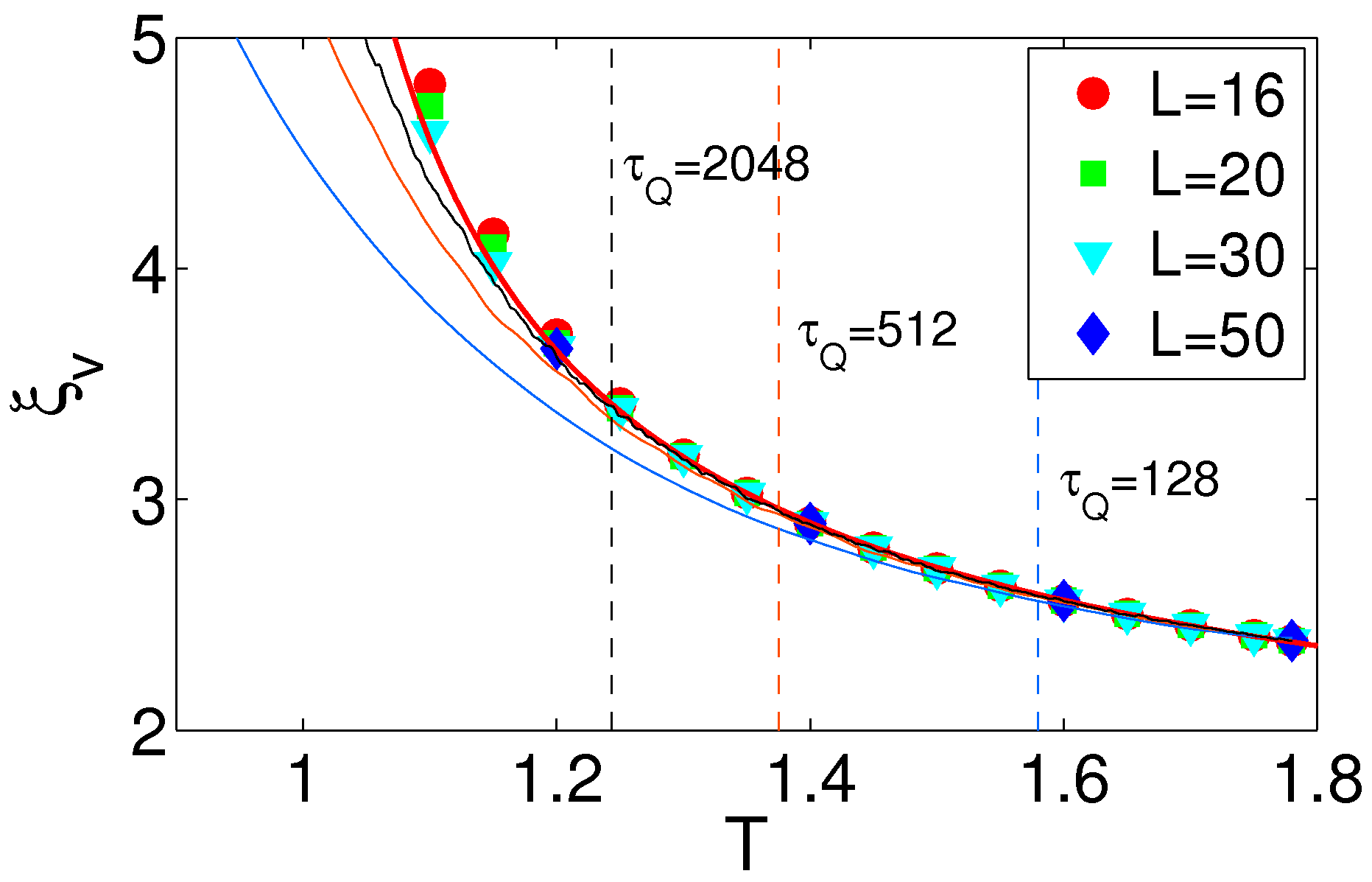}
\includegraphics[width=0.475\textwidth]{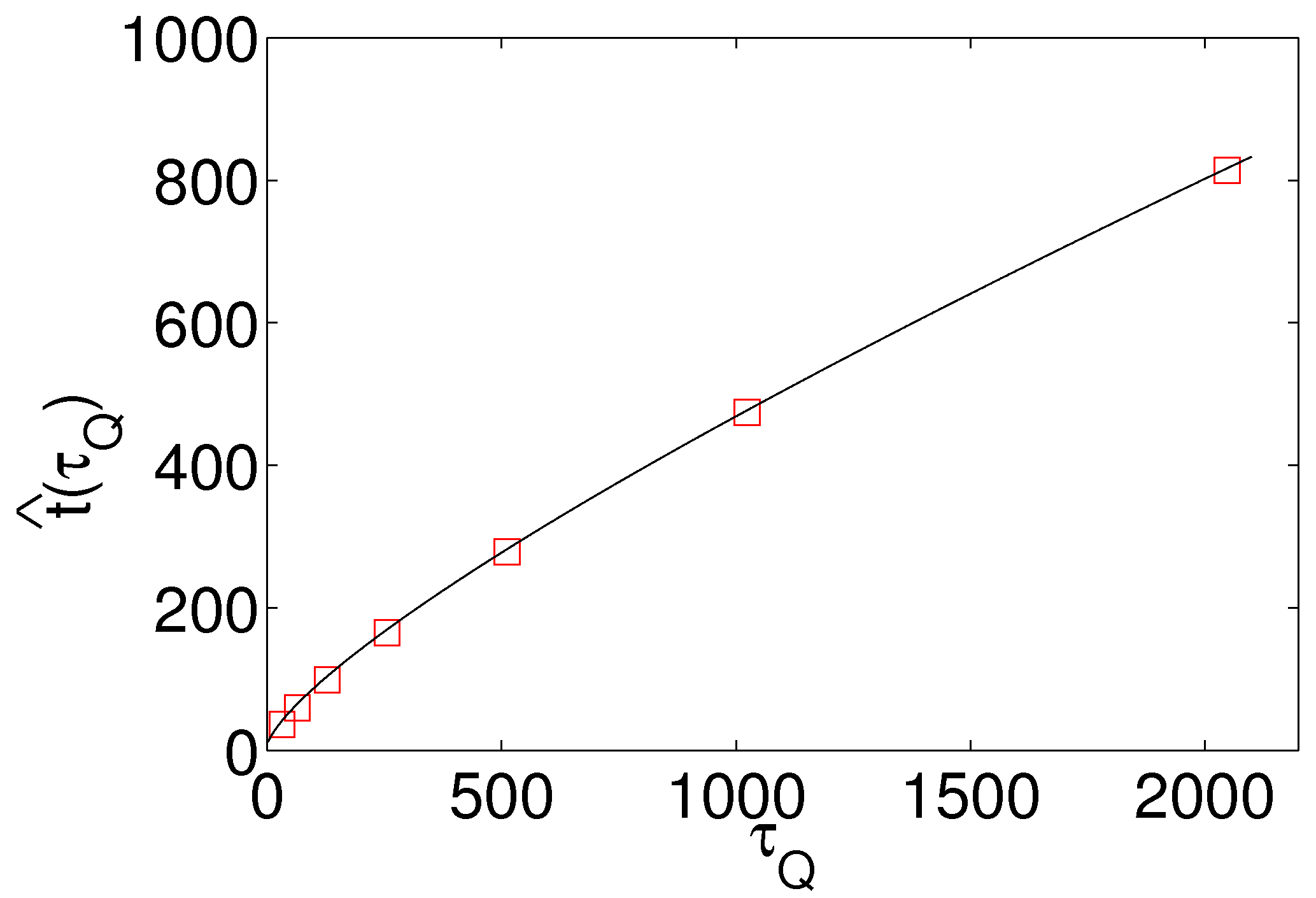}
\end{center}
\vspace{0.2cm}
  \caption{(Color online.) (a) The equilibrium correlation length,
    $\xi_v$, at $T>T_{KT}$, for different linear system sizes given in
    the keys are shown with points. The BKT exponential divergence, as
    obtained by fitting the data for temperatures higher than $T=1.2$,
    where we do not have strong finite size effects, are shown with
    think (red) lines.  The time-dependent length $\xi_{v}[T(t)]$
    for different cooling rates indicated on the figure are shown
    with thin lines. The predicted $\hat T$s, for different $\tau_Q$,
    are signaled with dashed vertical lines. See the text for more
    details. (b)~The crossover time $\hat t$ as a function of $\tau_Q$. The
  data points have been estimated from the numerical simulation -- see panel~(a) -- 
  and the line is the analytic
  prediction in eq.~(\ref{eq:log-corr}).  
   }
\label{fig:eq-corr-length}
\label{fig:hatt-tauQ}
\end{figure}

In order to put the prediction for $\hat \xi$ to the test we
compared the equilibrium and growing length extracted from the
numerical data.  We calculated the equilibrium values using systems
with small linear sizes, $L=16, 20, 30, 50$, that we let equilibrate.
In Fig.~\ref{fig:eq-corr-length}~(a) we display equilibrium data  for
 $\xi_v \propto \rho_v^{-1/2}$ with points. The
data for $T\geq 1.2$ do not show finite size effects and are in
agreement with the ones in~\cite{Gupta92} (obtained using much larger
samples, $L=512$) while at lower-$T$ the spread of data is severe and
the maximum values found for $L=50$ are way below the values estimated
in~\cite{Gupta92}.  
The BKT exponential divergence is reported on the
figure with solid line. 
The dynamic
data for $\xi_{v}[T(t)]$ obtained by using different cooling rates,
$\tau_Q=128, \ 512, \ 2048$ MCs, are shown with thin lines in both
panels.  Finally, the $\hat T$-values at which the dynamic data should
diverge from the static ones are signaled with dashed vertical lines.
They are obtained from the numerical solution of
eq.~(\ref{eq:cross-over}), by using the $b_{\xi}$ stemming from the
fit of the BKT exponential divergence. The parameter $a_{\tau}$ is, on
the other hand, determined in such a way that the resulting
$\hat{T}$-values give a good estimate of the divergence of the dynamic
data from the static ones.

In Fig.~\ref{fig:hatt-tauQ}~(b) the numerically determined 
$\hat{t}$ data are compared to the analytic estimate~(\ref{eq:log-corr}).
The agreement between Monte Carlo data and analytic estimate is very good 
within our numerical accuracy.

\subsection{Out of equilibrium evolution}

At time $-\hat t$ close to criticality, the system falls out of
equilibrium but it continues to evolve further, from $\hat T = T(-\hat
t) > T_{KT}$ to $T(\tau_Q) = 0$ following the out of equilibrium
critical dynamics rules.  In the case of a finite rate quench, the
total time spent in a critical region includes time $\hat t$, which
the system spent above $T_{KT}$ after falling out of equilibrium. That
is to say, the total time evolving with critical dynamics is $\Delta t
= \hat t + t$. Then, we propose to match the equilibrium relaxation
above $\hat T$ with the critical one below $\hat T$ with the
asymptotic scaling of the growing length
\begin{eqnarray}
\xi(t) = \left\{
\begin{array}{ll}
\xi_{eq}[T(t)]  \qquad \qquad \qquad\qquad 
& \qquad t < -\hat t(\tau_Q) 
\\
\hat\xi + \left\{ 
\lambda[T(t)] \
\frac{\Delta t}{\ln(\Delta t/t_0[T(t)])}
\right\}^{1/z}
\equiv \xi_{\rm low T}(t)
& \qquad t > -\hat t(\tau_Q) 
\label{eq:scaling-assumption}
\end{array}
\right.
\end{eqnarray}
Note that we assumed that the $T$-variation of the parameters
$\lambda$ and $t_0$ should be sufficiently slow so that we can simply
replace their time-dependent values in the growth-law after a quench.
We shall further fix $t_0=1$ as in the quenching case.
At $\Delta t=0$: $T(t)=T(-\hat t)=\hat T$, $\xi_{eq}(t) = \hat \xi$
and these formul\ae \ match.

\begin{figure}[h]
\hspace{1cm} (a) \hspace{5.5cm} (b)
\begin{center}
\includegraphics[width=0.45\textwidth]{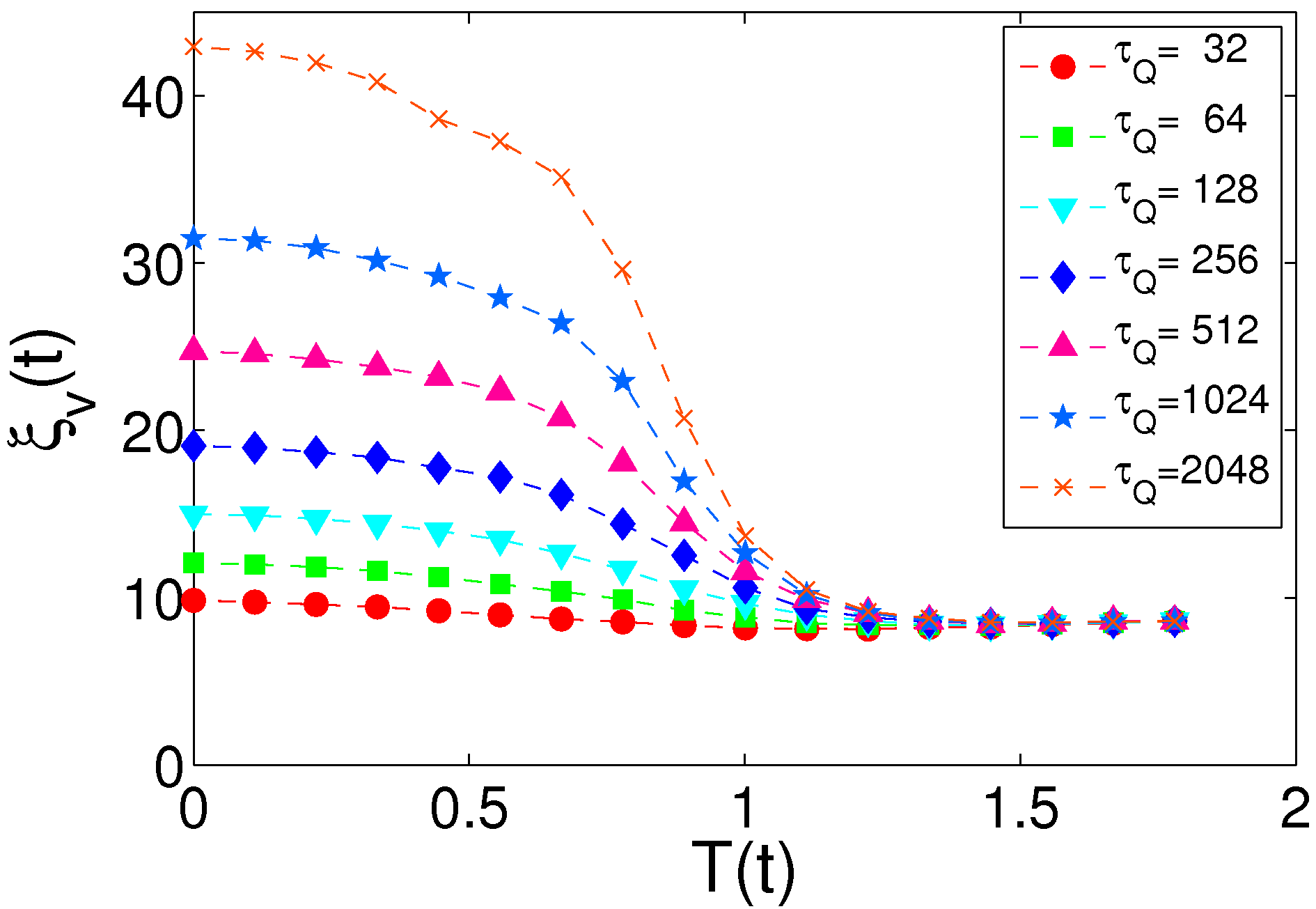}
\includegraphics[width=0.45\textwidth]{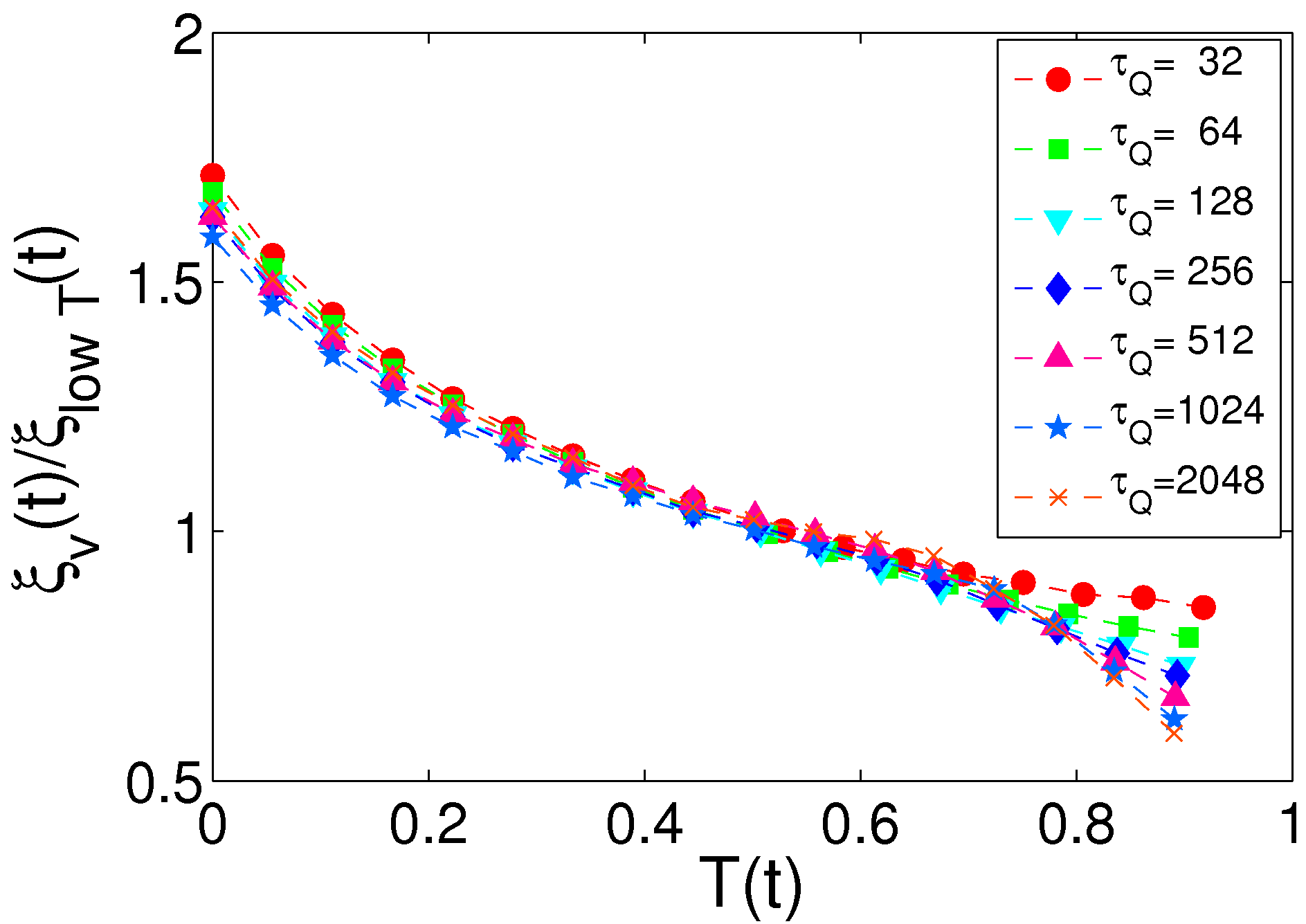}
\end{center}
  \caption{(Color online.)  Growing correlation length $\xi_v(t)$
    during annealing from $2T_{KT}$ to $T=0$ with different cooling
    rates $\tau_{Q}$. (a) Bare data for seven cooling rates given in
    the key.  (b) Rescaling with the proposed asymptotic
    form~(\ref{eq:scaling-assumption}) for $t > -\hat t(\tau_Q)$. }
\label{fig:annealing-scaling}
\end{figure}

Let us confront the order of magnitude of the two terms in the second
line of eq.~(\ref{eq:scaling-assumption}) for times of the order of
the inverse cooling rate, $t\simeq \tau_Q$.  Using Eq.~(\ref{eq:xi-hat}), 
the first term is $\hat \xi \simeq [\tau_Q/(\ln \tau_Q)^{(1+\nu)/\nu}]^{1/z}$.  Since
$\hat t$ is much smaller than $\tau_Q$, in the long $\tau_Q$ limit
$\Delta t\simeq \tau_Q$. The $\lambda$ prefactor is a function of $T$
and hence of $t$ but it should be bounded by the finite and
non-vanishing limiting values in Fig.~\ref{fig:T-dep}. Therefore,
apart from the finite prefactor originating in $\lambda$, the second
term is of the order $\simeq [\tau_Q/\ln \tau_Q]^{1/z}$.  Using
$\nu=1/2$, one concludes that the second term, describing the out of
equilibrium dynamics below $\hat T$, dominates in the large $\tau_Q$
limit.  (The same occurs in the conventional second order phase
transitions although the mechanism is different~\cite{Biroli}.)
This fact can be observed in panel~(a) of Fig.~\ref{fig:annealing-scaling}
where the growing $\xi_v$ under annealing with seven cooling rates 
given in the key is shown as a function of the time-varying temperature. 
 
In Fig.~\ref{fig:annealing-scaling} we present data for the dynamic
growing length for different cooling rates. Panel~(a) displays the
bare data for seven inverse cooling rates $\tau_Q$ given in the key.
Panel~(b) is scaling plot that tests the hypothesis in
Eq.~(\ref{eq:scaling-assumption}) in the low-temperature region during
the annealing process. We use the values of the parameter
$\lambda[T(t)]$ obtained from the linear fit of data for the quench in
Fig.~\ref{fig:T-dep}. Before rescaling, the correlation length
$\xi_{v}(t)$ grows with time, {\it i.e.},\ with decrease of
temperature, in such a way that at a given $T$, the value of $\xi_{v}$
increases with $\tau_{Q}$ (see panel~(a)).  After rescaling with the
proposed asymptotic form (\ref{eq:scaling-assumption}) for $t > -\hat
t(\tau_Q)$, the values of $\xi$ fall on top of each other. Perfect agreement 
with the analytic prediction would be the flat master curve  $\xi_v(t)/\xi_{\rm low T}(t)\equiv 1$.
The master curve deviates from this prediction but the range of variation 
is relatively small, varying from 1 to 1.6, {\it circa}. We ascribe this 
mismatch to potential non-linear corrections to 
$\lambda[T(t)]$ and small variations in $t_0[T(t)]$ that we have not taken into account.

\subsection{Numerical measurements of vortex density}

\begin{figure}[h]
 \begin{center}
\includegraphics[width=0.6\textwidth]{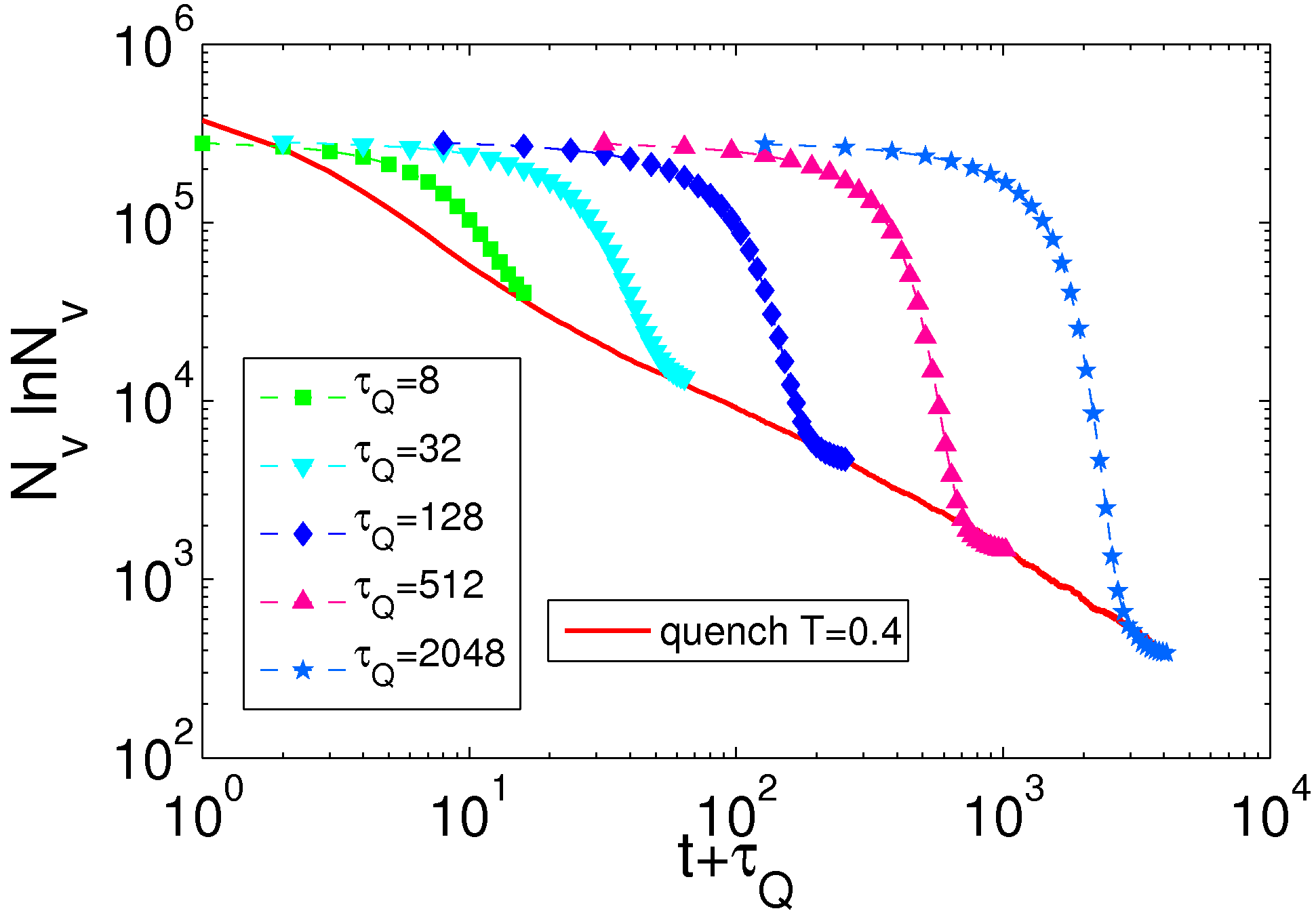}
\end{center}
  \caption{(Color online.) Time-dependence of the total number of vortices
    after annealing to $T=0$ with different inverse cooling rates $\tau_Q$
    given in the key.  The data are shown with points in the form
    $N_v \ln N_v$ as a function of $t+\tau_Q$. With a (red)  line we plot the
    number of vortices after an infinitely rapid quench to $T=0.4\leq T_{KT}$.}
\label{fig:rho-density-annealing}
\end{figure}

In Fig.~\ref{fig:rho-density-annealing} we study the density of all
vortices during annealing. We present the data obtained by
using five cooling rates in a log-log plot.  The translation of the
$x$ axis by $\tau_Q$ is done to give a common origin to all curves as
well as to allow for a direct comparison with the results
in~\cite{Chu}.

Four regimes are evidenced in the figure.  At very short times, as
compared to a function of $\tau_Q$, the data remain close to the
initial and rather large value $\rho_v\simeq 0.2$. This regime crosses
over to one with a rapid relaxation in which the density of vortices
decreases significantly (the shoulder in the curves). The crossover
occurs at $t \stackrel{>}{\sim} -\hat t$. The relaxation continues in
a smoother way and it is close to a power-law decay with logarithmic
corrections as expected. This regime ends when the curves reach the
limit of the instantaneously quenched data to $T=0.4$ that
they subsequently follow until $t=\tau_Q$ and $T=0$. In our setting
the density of defects {\it increases} with increasing $\tau_Q$ at
fixed $t+\tau_Q$.

The qualitative features of our data are very close to the ones
obtained by Chu and Williams~\cite{Chu} for a quench or annealing from
equilibrium at $T_{KT}$ to below the critical point. As already said,
their method is to solve numerically the Fokker-Planck equation for
vortex-pair dynamics in conjunction with the Kosterlitz-Thouless
recursion relations derived in~\cite{Ambegaokar}.  This approach
can only be used for initial conditions at and below $T_{KT}$ and does
not include the effect of free-vortices. In consequence, the scalings
found are consistent with power-laws without logarithmic corrections.

\begin{figure}[h]
\begin{center}
\includegraphics[width=0.6\textwidth]{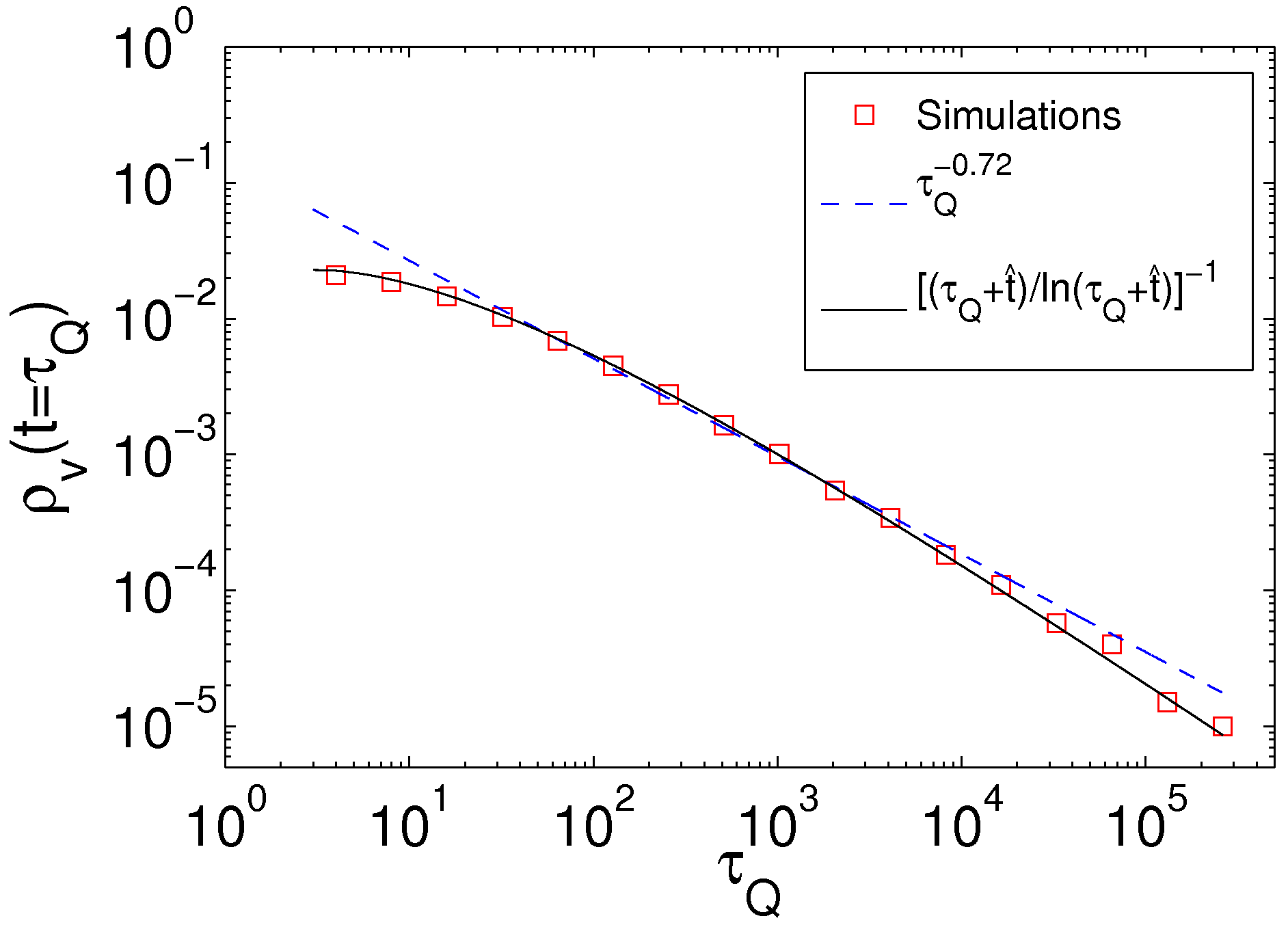}
\end{center}
\caption{(Color online.) $\tau_Q$-dependence of the density of
  vortices after annealing the system to $T=0$. Points represent the
  numerical data obtained from the direct counting of topological
  defects, the dashed (blue) line shows an effective power law
  $\tau_Q^{-0.72}$, while the full (black) line corresponds to the
  prediction in Eq.~(\ref{eq:rhov_T=0}). Similar results are obtained
  from $\rho_v \simeq \xi_2^{-2}$ (not shown).}
\label{fig:nv-tauQ}
\end{figure}

Finally, we extracted the remanent number of vortices at the end of
the cooling procedure, $t=\tau_Q$ and $T=0$, and we plot it as a
function of $\tau_Q$ in Fig.~\ref{fig:nv-tauQ}. The data
correspond to the direct counting of (free) defects.  The
total number of defects varies from $10^4$ to $10$ circa.
On the analytic side, 
the scaling assumption in Eq.~(\ref{eq:scaling-assumption}) suggest
\begin{eqnarray}\label{eq:rhov_T=0}
\rho_v(\tau_Q) &\simeq &
\tau_Q^{-1} 
+
\left\{
\lambda[T(\tau_Q)] \
\frac{\Delta t(\tau_Q)}{\ln(\Delta t(\tau_Q)/t_0[T(\tau_Q)])} 
\right\}^{-1} 
\nonumber\\
&=& 
\tau_Q^{-1} + \left\{ \lambda \
\frac{(\tau_Q+\hat t)}{\ln[(\tau_Q+\hat t)/t_0]} 
\right\}^{-1}
\nonumber\\
&\simeq &
 \left\{ \lambda \
\frac{(\tau_Q+\hat t)}{\ln[(\tau_Q+\hat t)/t_0]} 
\right\}^{-1}
\end{eqnarray}
where $\lambda$ and $t_0$ in the last expression are the
zero-temperature values.  We compared the above assumption with the
numerical data in Fig.~\ref{fig:nv-tauQ} by using the values of
$\hat{t}$ obtained numerically in Sect.~\ref{sec:crossover} and we
found good agreement within our numerical accuracy. The fact that the 
data points bend at short $\tau_Q$ is well captured by the addition of
$\hat t$ in the numerator. This term becomes negligible for longer
$\tau_Q$. At intermediate values of $\tau_Q$ the decay can be easily
confused with an effective power-law, $\tau_Q^{-0.72}$, shown in
the figure with a dashed (blue) line. For longer values of $\tau_Q$ the data confirm 
 the logarithmic correction.  Note that the decay is slower
than in the discrete symmetry breaking case with a low-$T$ phase with
trully long-range order, where one finds $\rho_v(t=\tau_Q) \simeq \tau_Q^{-1}$~\cite{Biroli}.

\section{Conclusions}
\label{sec:conclusions}

This paper should contribute to the understanding of the dynamics of 
topological defects across classical thermal phase transitions. 

The framework we focused on is one in which the system of interest is
coupled to an environment in equilibrium at a given temperature.  The
contact to the bath induces dissipative stochastic dynamics that, for
concreteness, we chose not to conserve the order parameter. The system
is taken across a phase transition by either varying the temperature
of the external bath or by tunning a parameter in its Hamiltonian.
The disordered (high temperature) phase is plagued with topological
defects. After entering the ordered (low temperature) phase the
system's configuration is out of equilibrium and the system relaxes by
changing their organization and reducing their density.  The nature
and structure of topological defects depends strongly on the dimension
of the order parameter, $n$, and the dimension of space, $d$.  For the
scalar case, $n=1$, the defects are domain walls and these are points
in $d=1$, lines in $d=2$ and surfaces in $d=3$.  For $n=d$ the defects
are point-like and the example at hand are vortices in $d=2$.  While
the ordering kinetics of systems subjected to instantaneous quenches
has been studied at length with field theoretical and numerical
techniques~\cite{coarsening-reviews}, the analysis of the effect of
infinitely slow cooling rates has not been developed to the same
extent
(see~\cite{Stinchcombe,FisherHusecooling,Yoshino,Krapivsky}). The
Kibble-Zurek mechanism -- based on critical scaling above the critical
point -- provides a concrete prediction for the density of defects
left over across a phase transition when this is crossed at very low
rate. This proposal captures correctly the moment and parameter value
at which the system falls out of equilibrium~(see~\cite{Krapivsky} for
a very detailed calculation in the $1d$ Glauber Ising chain) but
neglects, incorrectly in dissipative thermal phase transition, the
relaxation dynamics in the ordered phase.  In~\cite{Biroli} the case
of a second-order phase transition with discrete spontaneously broken
symmetry was discussed. In this paper we studied a different type of
transition, the BKT one in the $2d$ XY model.

To summarize our results, we first analyzed in more detail than
previously done the out of equilibrium relaxation of a $2d$ XY model
infinitely rapidly quenched from $T_0\to \infty$ to $T<T_{KT}$.  Using
MC simulations we confirmed the functional form of the growth law
$\xi(T,t) \simeq [\lambda(T) t/\ln t/t_0]^{1/2}$ derived
in~\cite{Huse,Bray94} and we determined its temperature dependence
numerically. Although the log-correction has a theoretic foundation,
this law has been frequently confused with an effective power law,
especially in the field of freely decaying $2d$ turbulence as modelled
with a Ginzburg-Landau approach~\cite{Huber}.  We also analyzed the
evolution of the structural properties such as the distribution of
pair distances or the polarity correlation function in the course of
time and we found that they all satisfy dynamic scaling with respect
to the same growing length. Importantly enough, we demonstrated that
the out of equilibrium dynamic properties are determined by the
density of free vortices in the low-$T$ phase as opposed to the
density of all vortices that includes the equilibrium contribution of
bound pairs. This fact is the reason for the failure of dynamic
scaling when the length determined from the decay of the total density
of vortices was used~\cite{Rojas99,Bongsoo}.  Next, we studied cooling
rate dependencies, we proposed a scaling form for the growing length
under these circumstances, and we checked it numerically. We found
that the density of free topological defects not only depends on the
cooling rate used but does also on the time spent in the low
temperature phase, where vortices and anti-vortices tend to bind or
annihilate. In particular, the density of vortices at $t\simeq \tau_Q$
is $\rho_v \simeq [\tau_Q/\ln\tau_Q]^{-1}$ for very long $\tau_Q$.
In short, the dynamics in the critical low-$T$ phase of the $2d$ XY is
crucial to determine the density of topological defects.

\vspace{1cm}

\noindent
\underline {Acknowledgements:} This research was financially supported
by ANR-BLAN-0346 (FAMOUS),
in part by the National
Science Foundation under Grant No. NSF PHY05-51164, and
in part by the Swiss National Science Foundation (SNSF).  We thank very
useful discussions with G. Biroli, P. Calabrese, D. Dom\'{\i}nguez,
G. Lozano, R. Rivers, and A. Sicilia.

\end{document}